\documentclass[]{aa}

\usepackage{graphicx} 
\usepackage{natbib}
\usepackage{txfonts}
\begin{document}

\title{Detection of steam in the circumstellar disk around a massive
  Young Stellar Object \thanks{Based on observations collected at the
    European Southern Observatory at La Silla and Paranal, Chile (ESO
    Programmes 68.C-0652 and 69.C-0448)}}

\author{W.-F. Thi \inst{1,2}
           \and
           A. Bik \inst{1,3}}

\institute{Sterrenkundig Instituut Anton Pannekoek,
           University of Amsterdam, Kruislaan 403 1098 SJ Amsterdam,
           the Netherlands \and ESA Research and Scientific Support Department, ESTEC, Keplerlaan 1, 2201 AZ, Noordwijk, the Netherlands \and European Southern Observatory, Karl-Schwarzschild-Strasse 2, 85748 Garching, Germany}

\offprints{thi@science.uva.nl}%

\titlerunning{Steam in the circumstellar disk around a massive YSO}

\date{Received ... ; Accepted ...}

\abstract{We report on the observation of hot water vapor (steam) in
  the inner AU of a young massive star located in the star-forming
  region \object{IRAS 08576-4334}. The water lines are detected in a
  medium resolution ($R\sim$10,000) $K$-band spectrum taken by the
  infrared spectrometer {\em ISAAC} mounted on the {\em VLT-ANTU}.
  The water vapor is at a mean temperature of 1565$\pm$510~K, cooler
  than the hot CO found in the same object, which is at $\simeq$
  1660~K and the column density is $N$(H$_2$O)$=$ (2.5 $\pm$ 0.4)
  $\times$ 10$^{18}$ cm$^{-2}$.  The profile of both H$_2$O and CO
  lines is best reproduced by the emission from a Keplerian disk.  To
  interpret the data, we also investigate the formation of molecules
  and especially CO and water vapor in the inner hot and dense part of
  disks around young high mass stars using a pseudo time-dependent
  gas-phase chemical model.  Molecules are rapidly photodissociated
  but this destruction is compensated by an efficient formation due to
  fast neutral-neutral reactions.  The ability of CO molecules to
  self-shield significantly enhances its abundance. Water molecules
  are sufficiently abundant to be detectable. The observed H$_2$O/CO
  ratio is reproduced by gas at 1600~K and an enhanced UV field over
  gas density ratio
  $I_{\mathrm{UV}}/n_{\mathrm{H}}$=10$^{-4}$--10$^{-6}$.  The
  simulations support the presence of CO and H$_2$O molecules in the
  inner disks around young massive stars despite the strong UV
  radiation and show that the OH radical plays an essential role in
  hot gas chemistry.  }

% CO overtone emission lines are only detectable at high column densities.

\maketitle
% ------------------------------------------------------------------
%
%
%
%
%
% ------------------------------------------------------------------
\section{Introduction}

Disks are found around many low-mass young stars and are natural
byproducts of the star formation process (e.g.,
\citealt{Waters1998ARA&A..36..233W};
\citealt{Natta2000prpl.conf..559N} for the Herbig~Ae stars and
\citealt{Mundy2000prpl.conf..355M} for the T~Tauri stars).
Unfortunately, observations of disks around their higher counterparts
remain inconclusive \citep{Churchwell2002ARA&A..40...27C}. The
presence or absence of a gaseous disk may help understanding the
detailed physics of high-mass star formation.  One of the main
difficulties in the formation of massive stars by direct molecular
core collapse is that the radiation pressure of a luminous massive
protostar will halt the accretion of matter, ultimately limiting the
final mass of the star \citep{Wolfire1987ApJ...319..850W}. Two major
scenarii have been proposed to overcome this shortcoming. In the first
one, massive stars can form by coagulation of two lower mass objects
in a dense star cluster \citep{Bonnell2002MNRAS.336..659B}.  In the
second scenario, a massive circumstellar disk will allow a significant
amount of matter to accrete to the star
(\citealt{Yorke2002ApJ...569..846Y};
\citealt{Jijina1996ApJ...462..874J}). Evidence for the presence of a
dense gaseous disk close to the star \object{NGC2024-IRS2} were
presented by \cite{Lenorzer2004A&A...414..245L}. The main argument is
that the infrared excess observed in \object{NGC2024-IRS2} can be best
explained by the emission from a dense gaseous disk extending up to
0.6~AU.  The knowledge of the structure and composition of the inner
disk is essential for our understanding of the accretion of matter
onto the star.  At the high densities and temperatures (100-2000~K)
characteristic of disks at 0.1-5~AU around young stars, molecules are
expected to be sufficiently excited to produce a rich ro-vibrational
spectrum in the near- and mid-infrared.  CO overtone emission, which
is most likely emitted by a disk, has been detected and analyzed by
\cite{Scoville83} toward \object{Orion BN object} and by
\cite{Chandler1995ApJ...446..793C} toward \object{NGC2024-IRS}.  Disk
models have been successful in modeling CO bandhead emission in
low-mass as well as high-mass young stellar objects
(\citealt{Najita1996ApJ...462..919N};
\citealt{Chandler1995ApJ...446..793C}; \citealt{Kraus00}). Cooler CO
fundamental emission, which probes the planet forming region of disks,
has been detected in a large number of low-mass pre-main-sequence
stars (\citealt{Blake2004ApJ...606L..73B};
\citealt{Najita2003ApJ...589..931N};
\citealt{Brittain2003ApJ...588..535B}).

Recent observations of CO overtone emission from young early type
stars further support the disk hypothesis (\citealt{COletter04};
  \citealt{Blum2004}).  The presence of CO molecules raises the
question of their formation and/or survival in the harsh environment
in the close vicinity of massive stars. Molecules other than CO are
also expected to exist in the inner few AU of disks. Carbon, nitrogen
and oxygen are the most abundant elements after Hydrogen and Helium.
At high temperatures and low pressures the dominant equilibrium
species of these elements are CO, N$_2$, and H$_2$O; at intermediate
temperatures, CH$_4$, N$_2$ and H$_2$O; and at low temperatures and
high pressures, CH$_4$, NH$_3$, and H$_2$O
\citep{Lewis1980ApJ...238..357L}. In all cases water vapor is the
second molecular reservoir of oxygen after CO in the gas phase at a
temperature between its dissociation temperature ($\sim$2500~K) and
its condensation temperature ($\sim$100-150~K).  Basically, all
gas-phase oxygen locked into H$_2$O does not participate to further
reactions (e.g., \citealt{Kaufman1996ApJ...456..611K};
\citealt{Elitzur1979ApJ...229..560E}). Chemical models of
protoplanetary disks abound in the literature for low-mass stars
(e.g., \citealt{Aikawa1999ApJ...519..705A};
\citealt{Markwick2002A&A...385..632M};
\citealt{Semenov2004A&A...417...93S}), high-mass stars (e.g.,
\citealt{Nguyen2002A&A...387.1083N}) or debris-disks (e.g.,
\citealt{Kamp2003A&A...397.1129K}) although most studies were limited
to the region located beyond a few AU. Nevertheless, several
  studies focus on the thermal-chemical structure of the inner disks
  around young low mass stars \citep{Glassgold2004ApJ...615..972G}.

Fundamental ro-vibrational and pure-rotational water lines have been
observed in the envelope of high-mass young stellar objects using the
Short- and Long-Wavelength Spectrometer on board the Infrared Space
Observatory (e.g., \citealt{Boonman2003A&A...406..937B};
\citealt{Wright2000A&A...358..689W};
\citealt{Helmich1996A&A...315L.173H}).  They find that freeze-out of
water onto grains predominates in the cold part of the envelope while
the high abundance of water in the warm part can only be explained by
sublimation of water molecules.

Water emission lines have also been observed in the sub-millimeter
domain with the Submillimeter Wave Astronomy Satellite (SWAS) but the
beam was too large to allow the study of small objects like the disks
around high mass young stellar objects.  Surprisingly, SWAS data show
that gaseous H$_2$O and O$_2$ are not the dominant carriers of
elemental oxygen in cold molecular clouds; the oxygen remains in
atomic form, providing strong constraints on chemical models
(\citealt{Melnick2000ApJ...539L..87M};
\citealt{Bergin2000ApJ...539L.129B}).

Highly excited water lines, which are absent in our atmosphere, are
detected from the ground in the atmosphere of M dwarfs (see
\citealt{Allard2001ApJ...556..357A} and references therein), Sunspots
\citep{Zobov2000ApJ...530..994Z} and in the disk around the low-mass
young star \object{SVS~13} \citep{Carr2004ApJ...603..213C}. In the
latter object, CO bandhead emission has also been detected.

In this paper, we model the hot water emission lines detected in the
disk around the massive young star \object{08576nr292} (\object{IRAS
  08576-4334}). The young stellar object (YSO) \object{08576nr292} is
one of the infrared sources detected in a survey of H{\sc ii} regions
by \cite{Kaper05}.  We also investigate the condition for the
formation and/or survival of molecules (mainly CO, H$_2$O) in a
gaseous disk around a massive young star. The inner disk around a
young massive star is probably too hot for dust grains to survive,
such that no H$_2$ formation on the grain surface can occur. Moreover,
far ultraviolet photons are no longer absorbed by dust so that
photodissociation occurs at high column densities.  In addition, the
disk temperature and density are high, making the chemistry different
compared to other environments.  For example, endothermic reactions or
reactions with an activation energy, mostly the neutral-neutral
reactions, and three-body reactions may dominate the chemical network
over the classical ion-molecule reactions. The chemistry scheme is
probably closer to that in shocks or protostellar winds than that of a
cold quiescent molecular cloud.

The paper is organized as follows: we first present the observations
and the fit to the $K$-band spectrum in Sect.~\ref{observations}. Then
we detail the chemical models in Sect.~\ref{chemical_models}.  In
Sect.~\ref{results}, we present and discuss the results of the
numerical runs. We finally conclude in Sect.~\ref{conclusion}.
  
% ------------------------------------------------------------------
\section{Observations}\label{observations}

\subsection{The high-mass young star \object{08576nr292}}

\object{IRAS 08576-4334}, identified by \cite{Liseau92} as IRS 34 in
their survey of IRAS sources in the Vela Molecular Cloud, is a
high-mass star forming region with total bolometric luminosity
$\log{(L/L_{\odot})}$= 4.35. The distance to the cloud has been
estimated to be around 700~pc. The extended H{\sc ii} region
\object{IRAS 08576-4334} includes a radio continuum source known as
\object{G265.151+01.454} \citep{Walsh99}.  \object{08576nr292} is one
of the near-IR sources detected in the $J$- and $K$-band image of
\object{IRAS 08576-4334} by \cite{Kaper05} and discussed in
\cite{Brgspec04}. The properties of \object{08576nr292} are listed in
Table~\ref{table_object_properties}.  After correction by 12 magnitude
of extinction, as derived by \citet{Brgspec04}, the $J_0-K_0$ color of
\object{08576nr292} is small (0.4). Because the intrinsic color of hot
stars is $\sim$-0.2, the near-IR excess from \object{08576nr292} is
$(J-K)$=0.6, suggesting no strong hot dust emission.

% -------------------------------------------------------------------
\begin{table}[ht] 
\begin{center}
\begin{tabular}{ll}
\hline\hline
\noalign{\smallskip}
\multicolumn{2}{c}{\object{08576nr292}}\\
\hline
\noalign{\smallskip}
Ra ($J$2000) & \phantom{-}08:59:21.6\\
Dec ($J$2000) & -43:45:31.6\\
$K$ (Mag) & \phantom{-}9.4 $\pm$ 0.01\\
$J-K$ (Mag) & \phantom{-}4.7 $\pm$ 0.09\\
$A_{\mathrm{V}}$ & 12\\
$K_0$ (Mag) & -3.1\\
$J_0-K_0$ (Mag) & \phantom{-}0.4\\
$d$(kpc) & \phantom{-}0.7\\
$M_*$($M_{\odot}$) & \phantom{-}6\\
\noalign{\smallskip}
\hline
\end{tabular}  
\caption{Properties of \object{08576nr292}. The position and observed
  magnitude and color is taken from \citet{Kaper05}. The extinction
  estimate and the derived unreddened magnitude and color are taken
  from \citet{Brgspec04}.\label{table_object_properties}}
\end{center}
\end{table}
% -------------------------------------------------------------------

\subsection{Observation \& data reduction}

The spectrum of \object{08576nr292}, identified in the line of sight
to \object{IRAS 08576-4334}, was observed with ISAAC at the {\em Very
  Large Telescope} ANTU in 2002 at resolving power of 10,000 (slit
width of 0.3\arcsec). The near-infrared data were reduced in a
standard way using a combination of the dedicated software package
Eclipse and IRAF. We made use of flatfield taken during the night and
arc frames were taken by the {\em European Southern Observatory} (ESO)
staff during the day.

Standard stars of spectral type A observed at similar airmass were
used to correct for telluric OH emission and absorption. The
wavelength calibration was achieved by using arc lines in combination
with atmospheric OH lines. The details of the observations and
datareduction are presented in \cite{Brgspec04}.  The chosen setting
covers the spectral range from 2.280 till 2.400 micron so that the
four first CO bandhead ($\Delta v$=2, where $v$ is the vibrational
quantum number) can be observed. At resolving power 10,000 ($\Delta v
\simeq$ 30 km s$^{-1}$), the profile of the lines is not resolved and
a few lines may even be blended.
% -------------------------------------------------------------------
   \begin{figure}[ht]
     \centering
     \resizebox{\hsize}{!}{\includegraphics[]{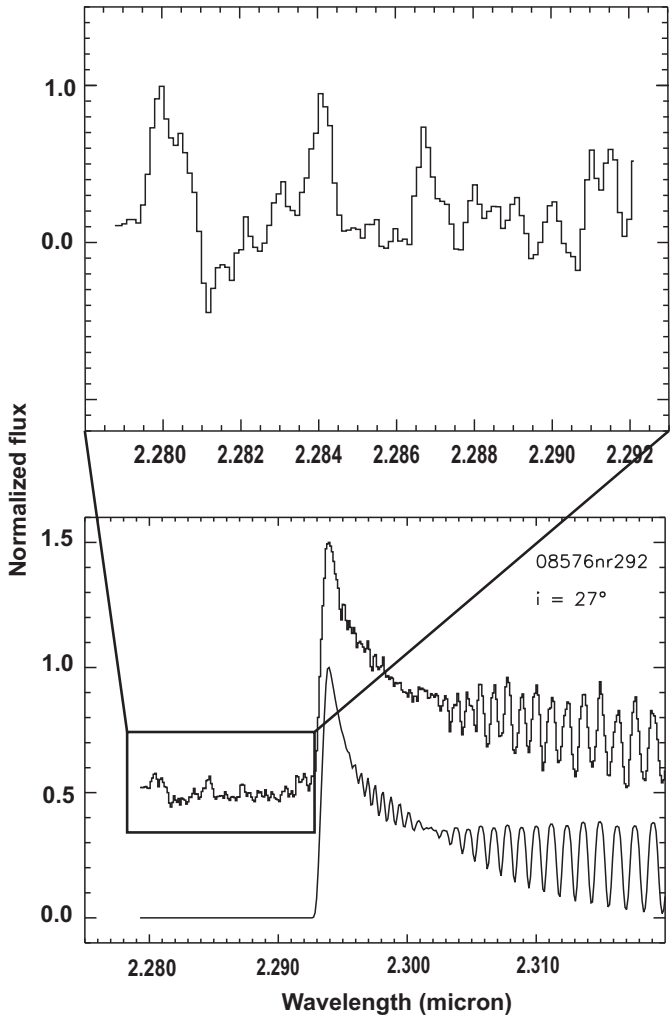}}
   \caption{The lower panel shows the best fit to the CO first bandhead toward
     \object{08576nr292}. In the upper part the observed spectrum is
     shown. The spectrum is continuum subtracted and normalized to the
     peak of the 2--0 bandhead. The best fit of the CO-bandhead is
     plotted in the lower part. The parameters of the fit are
     discussed in \cite{COletter04}. Features shortward of 2.2935
     $\mu$m are shown in detail in the upper panel.\label{Fig_COfit}}
\end{figure}
% ----------------------------------------------------------------

\begin{figure}[ht]
  \centering
  \resizebox{\hsize}{!}{\includegraphics[]{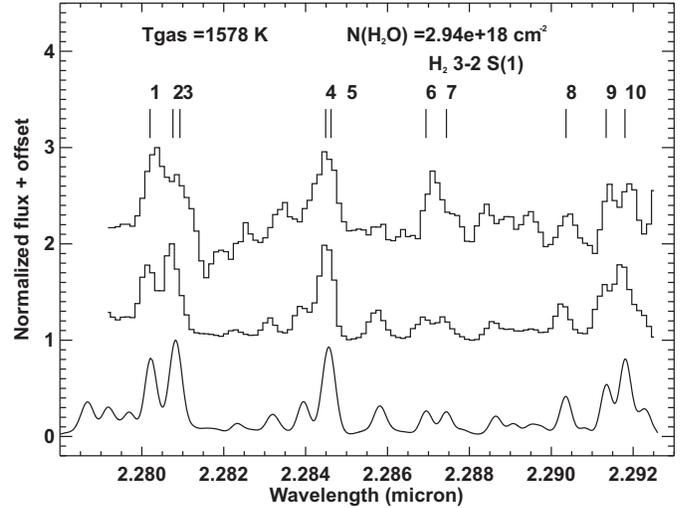}}
\caption{Best fit to the hot water lines with the slab model. 
  The upper spectrum is a blowup of the spectrum showed in Fig.
  \ref{Fig_COfit} between 2.278 and 2.292 $\mu$m, normalized to the
  maximum strength at $\simeq$2.280 $\mu$m.  The middle one represents
  the best fit while the lower curve is the synthetic spectrum before
  degradation at the resolution of the observed spectrum. The
  characteristic of the strong detected lines, which are labeled 1 to
  10, are given in Table~\ref{table_strongest_lines}. The H$_2$ 3-2
  S(1) emission line is located at position 6.\label{Fig_H2Ofit}}
\end{figure}
% ----------------------------------------------------------------

\subsection{Observed water lines}

% -------------------------------------------------------------------
\begin{table*}[ht] 
\begin{center}
\caption{Strongest detected water emission lines. Various spectroscopic characteristics of the detected lines are also provided.
  The last column lists estimates of the critical density
  $n_{\mathrm{crit}}$ assuming $T_{\mathrm{gas}}$=1000~K and $\sigma
  \simeq$10$^{-15}$ cm$^{-2}$.
  \label{table_strongest_lines}}
\begin{tabular}{llllllll}
\hline\hline
\noalign{\smallskip}
\multicolumn{1}{c}{Line}& \multicolumn{1}{c}{Nuclear state} &\multicolumn{1}{c}{Transition(u$\rightarrow$l)} & \multicolumn{1}{c}{Wavelength} & \multicolumn{1}{c}{$E_{\mathrm{upper}}$} & \multicolumn{1}{c}{$E_{\mathrm{lower}}$}&\multicolumn{1}{c}{$A_{\mathrm{ul}}$}&\multicolumn{1}{c}{$n_{\mathrm{crit}}$}\\
 & & \multicolumn{1}{c}{$(v_1 v_2 v_3)\ J\ K_a K_c\ -\ (v_1' v_2' v_3')\ J'\ K_a' K_c'$} & \multicolumn{1}{c}{(micron)} & \multicolumn{1}{c}{(cm$^{-1}$)} & \multicolumn{1}{c}{(cm$^{-1}$)} & \multicolumn{1}{c}{(s$^{-1}$)}&\multicolumn{1}{c}{(cm$^{-3}$)} \\
\hline
\noalign{\smallskip}
1....... & ortho-H$_2$O & \ \ \ \ (001) 15 \ 6 10           -- (000) 14 \ 4 11           & 2.28020     & 7131.63 & 2746.02 & 9.50 10$^{-2}$    & 9.50 10$^{8}$\\
2....... & ortho-H$_2$O & \ \ \ \ (001) 15 \ 7 \phantom{1}9 -- (000) 14 \ 5 10           & 2.28076$^{(1)}$ & 7302.72 & 2918.24 & 8.88 10$^{-2}$& 8.88 10$^{8}$\\
3....... & ortho-H$_2$O & \ \ \ \ (001) 16 \ 5 11           -- (000) 15 \ 3 12           & 2.28093$^{(1)}$ & 7464.48 & 3080.18 & 8.27 10$^{-2}$& 8.27 10$^{8}$\\
4....... & ortho-H$_2$O & \ \ \ \ (001) 14 \ 8 \phantom{1}6 -- (000) 13 \ 6 \phantom{1}7 & 2.28449$^{(1)}$ & 7133.78 & 2756.41 & 6.56 10$^{-2}$& 6.56 10$^{8}$\\
5....... & ortho-H$_2$O & \ \ \ \ (100) 18 \ 7 12           -- (000) 17 \ 4 13           & 2.28462$^{(1)}$ & 8394.96 & 4017.90 & 2.00 10$^{-1}$& 2.00 10$^{9}$\\
6....... & ortho-H$_2$O & \ \ \ \ (001) 12 \ 9 \phantom{1}3 -- (000) 11 \ 7 \phantom{1}4 & 2.28694    & 6694.58 & 2321.90 & 1.88 10$^{-2}$    & 1.88 10$^{8}$\\
7....... & ortho-H$_2$O & \ \ \ \ (001) 14 \ 5 10           -- (000) 13 \ 3 11           & 2.28744     & 6619.79 & 2248.06 & 6.24 10$^{-2}$    & 6.24 10$^{8}$\\
8....... & ortho-H$_2$O & \ \ \ \ (001) 13 \ 8 \phantom{1}6 -- (000) 12 \ 6 \phantom{1}7 & 2.29036     & 6799.95 & 2433.80 & 4.65 10$^{-2}$    & 4.65 10$^{8}$\\
9....... & ortho-H$_2$O & \ \ \ \ (001) 13 \ 4 10           -- (000) 12 \ 2 11           & 2.29134     & 6139.03 & 1774.75 & 3.20 10$^{-2}$    & 3.20 10$^{8}$\\
10.....  & ortho-H$_2$O & \ \ \ \ (001) 18 \ 7 11           -- (000) 17 \ 5 12           & 2.29180     & 8537.65 & 4174.38 & 3.21 10$^{-1}$    & 3.21 10$^{9}$\\ 
\noalign{\smallskip}
\hline
\end{tabular}  
\end{center}
\begin{flushleft}
{\em Note} $^1$ Blended lines.
\end{flushleft}
\end{table*}
% -------------------------------------------------------------------
The spectrum shows four prominent CO bandhead emissions which are
modeled in \cite{COletter04}. The fit to the first CO bandhead is
shown in the lower panel of Fig.~\ref{Fig_COfit}. The upper panel
displays in detail the spectrum shortward of $\lambda <$ 2.2935
$\mu$m. The CO bandhead is best fitted by gas at 1660~K located in a
Keplerian rotating disk extending from 0.2 till 3.6 AU and viewed with
an angle of 20--30$\degr$ with respect to the disk rotation axis. The
column density of CO is relatively large ($N$(CO)$\simeq$3.9 $\times$
10$^{21}$ cm$^{-3}$).

Several emission features can be seen blueward of the first CO
bandhead (i.e. $\lambda <$ 2.2935 $\mu$m), which we attribute to
  water vapor at T$>$1000~K. The emission features were checked
  against atmospheric transmission spectra to ensure no artifact is
  present.  Using spectroscopic parameters drawn from the
compilations HITEMP \citep{Rothman2003JQSRT..82....5R} and that of
\cite{Tennyson2001JPCRD..30..735T}, most lines (often blended) in the
2.280 -- 2.2935 $\mu$m region can be assigned to hot water emission
lines. HITEMP ({\em High-Temperature} molecular spectroscopic
database) contains water vapor spectroscopic parameters of lines which
have strength greater than 3 $\times$ 10$^{-27}$ cm$^{-1}$/(molecule
cm$^{-2}$) at 1000~K while the energy of the levels in
\cite{Tennyson2001JPCRD..30..735T} were carefully assessed using
energy levels derived from sophisticated variational methods. The
HITEMP database contains in total 1816 water lines between 2.278 and
2.292 $\mu$m which are sufficiently strong at
$T_{\mathrm{gas}}$=1500~K, but only 10 lines have strength greater
than 10$^{-23}$ cm$^{-1}$/(molecule cm$^{-2}$), which is likely the
threshold for visual detection in the observed spectrum. The detected
strong water lines and their spectroscopic characteristics are
summarized in Table \ref{table_strongest_lines}. The wavelength of
three of the features (lines 8, 9 and 10) matches those of hot water
lines detected by \cite{Carr2004ApJ...603..213C} in the inner disk
around the low mass young star \object{SVS~13}. The spectrum of
  \object{SVS~13} obtained by \cite{Carr2004ApJ...603..213C} is at a
  higher resolution but covers a much smaller spectral range.

It should be noticed that all 1816 lines are included in the synthetic
spectra described in Sect.~\ref{Spectral_modeling}. The position of
the lines is indicated in Fig.~\ref{Fig_H2Ofit} and
Fig.~\ref{Fig_disk_fit}.

All {\em strong} detected lines arise from the first vibrational
excited symmetric (100) and antisymmetric (001) stretch level of
ortho-H$_2$O.  It is known that there is a near coincidence of the
frequencies of those two modes. The rotational levels are high
($12<J<18$) as expected since only high level lines are not present in
our atmosphere and can be readily detected with ground telescopes. At
the resolving power of 10,000 (or resolution $\Delta v \simeq 30$ km
s$^{-1}$), the lines are unresolved and few of them blended.  The
observed lines arise from levels with energy from 5000-9000~K. The
detected (001)-(000) transitions have $\Delta K_{\mathrm{a}}=-2$ and
$\Delta K_{\mathrm{c}}=+1$ and are therefore of $a$-type. The
(100)-(000) transition is of $b$-type ($\Delta K_{\mathrm{a}}=-3$ and
$\Delta K_{\mathrm{c}}=+1$) (see Table~\ref{table_selection_rules}).

Water lines are clearly seen in this object because they are not
blended. The profile of emission lines coming from the inner part ($<$
a few AU) of a disk in Keplerian rotation can have $FWHM$ up to
$\simeq$ 200-250 km s$^{-1}$ (or $\Delta \lambda = 0.0015~\mu$m) if
viewed edge-on. In comparison water lines wavelength positions are
often separated by less than 65 km s$^{-1}$ (or 0.0005 $\mu$m).
Therefore the rotational broadening of all lines produces a blend of
features that can be mistaken for a continuum.

\subsection{Spectral modeling}\label{Spectral_modeling}

In this section we apply two different models to fit the observed
spectrum. The first model is a simplified version of the disk model
applied to the CO bandheads of this object by \citet{COletter04}. In
this model we assume that the gas is originating from an isothermal
slab of gas. The second model which is used to fit the data is similar
to the model used to fit the CO bandheads. The gas is located in a
Keplerian rotating, isothermal disk. First we describe the
ro-vibrational transitions of water vapor and the transition selection
rules.

\subsubsection{H$_2$O ro-vibrational transitions}

The structure of the water molecule is described in molecular physics
textbooks (e.g., \citealt{Bernath_book1995}). As a triatomic
asymmetric (near-oblate) top molecule, water has three vibrational
modes and three rotational modes. The vibrational levels are
characterized by three vibrational quantum numbers $v_1$ (symmetric
stretching mode $\nu_1$), $v_2$ (bending mode $\nu_2$) and $v_3$
(asymmetric stretching mode $\nu_3$). The energy of the lowest
vibrational levels is given in Table~\ref{table_h2o_energy_levels}.
The rotational levels are labeled using the standard asymmetric top
notation $J_{K_a K_c}$, where $J$ is the main rotational quantum
number, $K_a$ is the prolate quantum number (projection of $J$ along
the a axis) and $K_c$ the oblate quantum number (projection of $J$
along the c axis). Notice that $K_a+K_c=J$ or $J+1$. The rotational
statistical weight is ($g_J=2J+1$). The total degeneracy is
  $g=g_I g_J$.

Like H$_2$, H$_2$O has two hydrogen atoms each having a nuclear spin
1/2, which combine to a singlet (para-H$_2$O) and triplet
(ortho-H$_2$O).  Transitions between para and ortho states are highly
forbidden.  Ortho states have a nuclear statistical weight $g_I$ of 3
while para states have $g_I$ equal to 1. The sum $K_a+K_c+v_3$ is odd
for ortho states and even for para states.
% -------------------------------------------------------------------
\begin{table}[ht]
\begin{center}
\caption{Lower energy levels of vibration of water molecule in cm$^{-1}$
  with ground energy level at 0 cm$^{-1}$. The zero-point energy is
  4500 cm$^{-1}$. To convert the energy scale into K, multiply by
  1.438768.
\label{table_h2o_energy_levels}
}
\begin{tabular}{llll}
\hline\hline
\noalign{\smallskip}
\multicolumn{1}{c}{Symmetry} & \multicolumn{1}{c}{Vibrational level} & \multicolumn{1}{c}{Energy} & \multicolumn{1}{c}{Level type}\\
\multicolumn{1}{c}{}&\multicolumn{1}{c}{($v_1 v_2 v_3$)}   & \multicolumn{1}{c}{(cm$^{-1}$)} & \\
\hline
\noalign{\smallskip}
A$_1$ & 000 & 0000.00 & ground\\
A$_1$ & 010 & 1594.75 & fundamental \\
A$_1$ & 020 & 3151.63 & overtone \\
A$_1$ & 100 & 3657.05 & fundamental\\
B$_2$ & 001 & 3755.92 & fundamental\\
A$_1$ & 030 & 4666.79 & overtone \\
A$_1$ & 110 & 5234.97 & combination \\
B$_2$ & 011 & 5331.27 & combination \\ 
A$_1$ & 040 & 6134.01 & overtone \\
A$_1$ & 120 & 6775.09 & combination \\
B$_2$ & 021 & 6871.52 & combination \\
A$_1$ & 200 & 7201.54 & overtone \\
B$_2$ & 101 & 7249.81 & combination\\
A$_1$ & 002 & 7445.04 & overtone \\
\noalign{\smallskip}
\hline
 \end{tabular}  
\end{center}
\end{table}
% -------------------------------------------------------------------

The selection rules for asymmetric tops depend on the components
($\mu_{\mathrm{a}}$, $\mu_{\mathrm{b}}$, $\mu_{\mathrm{c}}$) of the
permanent dipole moment vector, along the a, b and c principal
molecular axes. The vibrational selection rules are $\Delta v =1$ for
the strong fundamental transitions and $\Delta v = 2$, 3, 4, ... for
the weaker overtone transitions. The rotational selection rules can be
divided into three general cases given in
Table~\ref{table_selection_rules} in addition to $\Delta J = 0, \pm
1$.

% -------------------------------------------------------------------
\begin{table}[ht]
\begin{center}
\caption{Selection rules for the ro-vibration transitions in addition to $\Delta J = 0,\ \pm1$. The transitions in brackets are weaker than the main transitions. For ro-vibrational transitions, the parity of $K_a+K_c+v_3$ has to be
  conserved. Transitions of type a only occur for oblate molecules
  whereas transitions of type c are only found for prolate molecules.
  Both oblate and prolate molecules can have b-type transitions.
\label{table_selection_rules}}
\begin{tabular}{llll}
 \hline\hline
 \noalign{\smallskip}
 \multicolumn{1}{c}{Transition}&\multicolumn{1}{c}{Criterion} & \multicolumn{1}{c}{$\Delta K_{\mathrm{a}}$}& \multicolumn{1}{c}{$\Delta K_{\mathrm{c}}$}\\
 \hline
 \noalign{\smallskip}
 $a$-type & $\mu_a \neq 0$ & \phantom{$\pm$}$0, \pm 2\ (\pm 4,\ ...)$    & $\pm 1\ (\pm 3,\pm 5,\ ...)$ \\
 $b$-type & $\mu_b \neq 0$ & $\pm 1, \pm 3\ ( \pm 5,\  ...)$   & $\pm 1, \pm 3\ ( \pm 5,\  ...)$\\
 $c$-type & $\mu_c \neq 0$ & $\pm 1\ (\pm 3,\pm 5,\ ...)$ & \phantom{$\pm$}$0, \pm 2\ (\pm 4,\ ...)$ \\
 \noalign{\smallskip}
 \hline
\end{tabular}  

\end{center}
\end{table}
% -------------------------------------------------------------------

\subsubsection{Slab model}

For the spectral modeling of the water lines we employ a method
similar to the fit of the CO bandhead emission.  The CO bandhead
emission is best modeled by a Keplerian disk seen at low inclination
with respect to the rotation axis.  First we further simplify the
model and assume that the water vapor lines arise from a isothermal
slab of gas at temperature $T_{\mathrm{gas}}$ with a column density
$N(\mathrm{H_2O})$. The remaining free parameter is the turbulent
velocity width $dv$. Values for $dv$ range from 5 to 25 km s$^{-1}$.
One should keep in mind that the profiles are unresolved and thus $dv$
is an ill-constrained parameter. The turbulent profile is a Gaussian
function. For $dv>$7 km s$^{-1}$, the gas is highly turbulent.
The transition data are taken from the compilation HITEMP.  The level
population is assumed to be at local thermodynamical equilibrium
(LTE).

An order of magnitude check of this last assumption can be performed.
The collisional de-excitation coefficients for the transition
u$\rightarrow$l is given by $\gamma_{\mathrm{ul}} \simeq
\sigma_{\mathrm{ul}} v_{\mathrm{rel}}$, where $\sigma_{\mathrm{ul}}$
is the de-excitation cross section and
$v_{\mathrm{rel}}=\sqrt{8kT_{\mathrm{gas}}/\pi m_{\mathrm{H_2O}}} =
10^5 \sqrt{T_{\mathrm{gas}}/1000}$ cm s$^{-1}$ is the mean relative
velocity of H$_2$O molecules. In the absence of published
cross-sections for the transitions considered here, we assume that
$\sigma_{\mathrm{ul}}$ is close to the geometrical cross section
$\sigma \sim$ 10$^{-15}$ cm$^{2}$ of the molecule. The rates of
spontaneous emission can be estimated from the Einstein coefficients
$A_{\mathrm{ul}}$, which range from 10$^{-5}$ s$^{-1}$ till 3 $\times$
10$^{2}$ s$^{-1}$ in the $K$-band region. The critical densities are
approximately $n_\mathrm{crit}=A_{\mathrm{ul}}/\gamma_{\mathrm{ul}}
\simeq (10^5 - 3 \times 10^{12}) \times
(T_{\mathrm{gas}}/1000)^{-1/2}$ cm$^{-3}$. The number density in the
inner part of disks is greater than 10$^{10}$ cm$^{-3}$ from the fit
to the CO bandhead by a LTE model. Therefore only the strongest lines
in the $K$-band are in non-LTE. More specifically, the lines detected
here have moderately strong Einstein coefficients
(10$^{-2}$--10$^{-1}$ s$^{-1}$) and by consequence relatively low
critical densities $n_{\mathrm{crit}}$ which are reported in the last
column of Table~\ref{table_strongest_lines}.
% ----------------------------------------------------------------

\begin{figure}[ht]
\centering
\resizebox{\hsize}{!}{\includegraphics[angle=90]{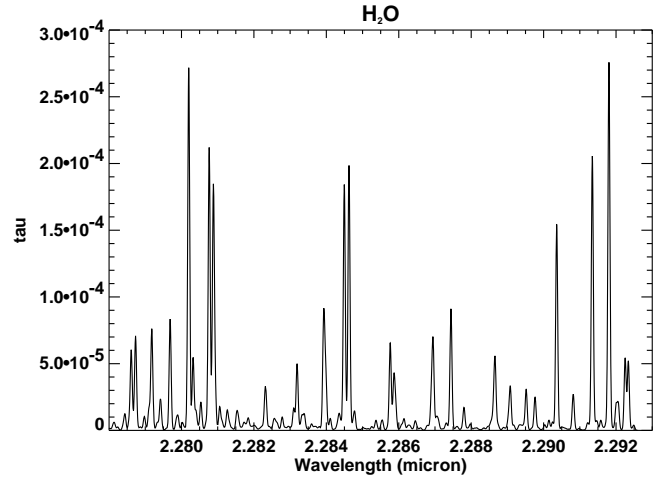}}
\caption{Typical line optical depth for the hot water lines with $dv$=5 km s$^{-1}$,
  $T_\mathrm{gas}$=1500~K and $N$(CO)=3 $\times$ 10$^{18}$ cm$^{-2}$.
\label{Fig_H2Otau}}
\end{figure}
% ----------------------------------------------------------------

The best model fit to the data was obtained by minimizing the $\chi^2$
using a Simplex \citep{Press1992nrfa.book.....P} and a genetic
algorithm methods ({\em Pikaia},
\citealt{Charbonneau1995ApJS..101..309C}) . The results of the two
methods were cross-checked to ensure that no local minima were
reached. The best fit using a slab model gives H$_2$O gas at 1550
  $\pm$ 510~K and column density of (2.9 $\pm$ 0.4) $\times$ 10$^{18}$
  cm$^{-2}$.  The best fit parameters are gathered in
  Table~\ref{tab_disk_model}.  Although the errors on the gas
  temperature and column density are dominated by systematics, we
  estimated the uncertainties in the two parameters following the
  method described in \cite{Bevington2003drea.book.....B}. The derived
  errors represent here an increase of $\chi^2$ by 1 from its value at
  the minimum. Noteworthy, the statistical significance of the errors
  is difficult to asses. The column density of water is 3 orders of
magnitude lower than CO while the temperature is only slightly lower
for water than CO. The intrinsic turbulent width $dv$ is an
ill-constrained parameter although a value of 5 km s$^{-1}$ has been
found to provide the best results. On the other hand the column
density and gas temperature are well constrained.
% -------------------------------------------------------------------
   \begin{figure*}[ht]
     \centering
     \resizebox{\hsize}{!}{\includegraphics[]{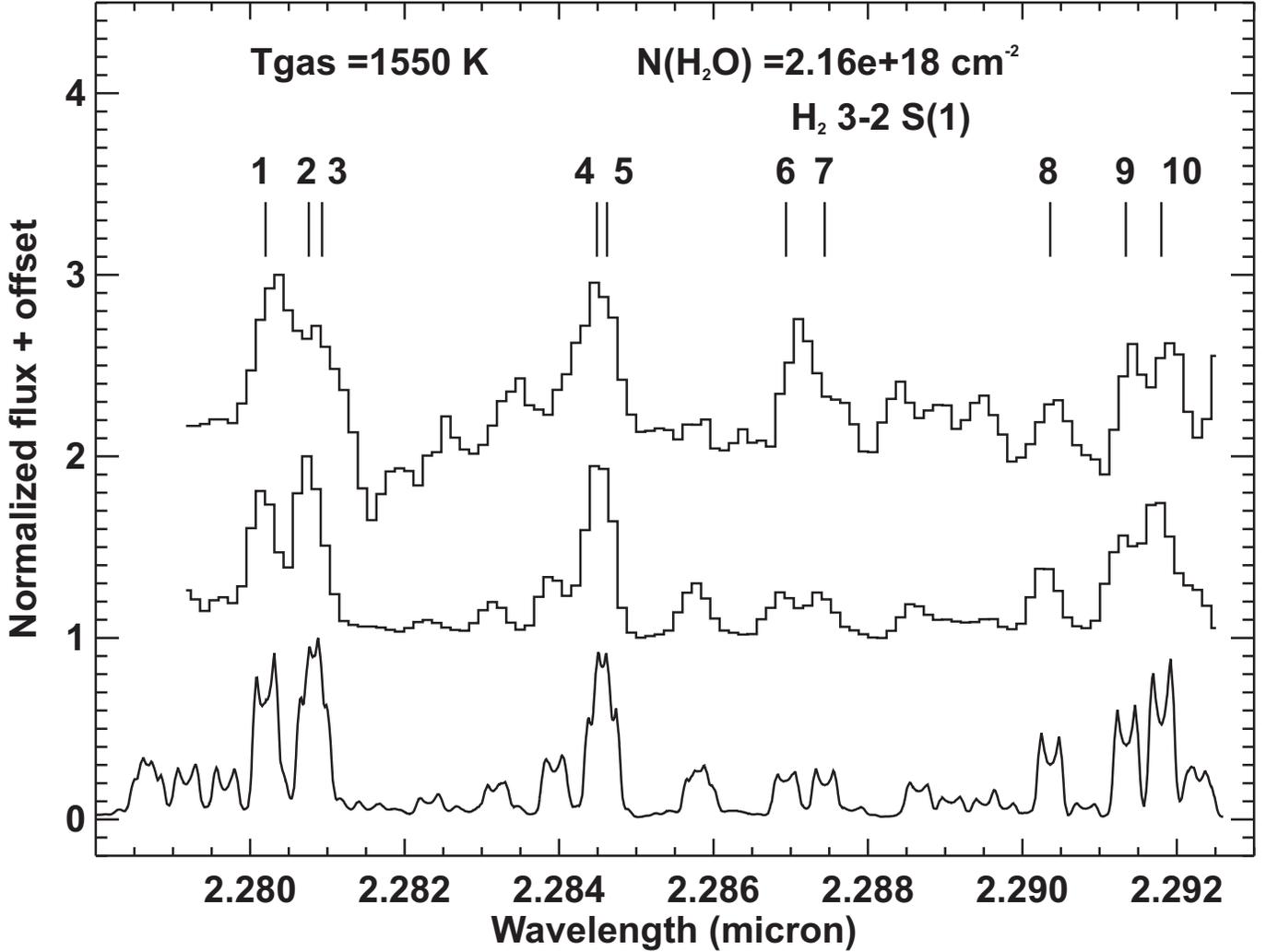}}
  \caption{Best fit to the water lines toward  \object{08576nr292} using the disk model. The upper curve
    is the observed normalized and continuum subtracted spectrum. The
    middle one represents the best fit while the lower curve is the
    synthetic spectrum before degradation at the resolution of the
    observed spectrum.  Notice the double-peak profile for lines which
    are not blended.
    \label{Fig_disk_fit}}
  \end{figure*}
% -----------------------------------------------------------------
\begin{table*}[ht]
\begin{center}
\caption{Parameters derived from the fit by the slab and disk model. The parameters for the CO bandhead fitting are derived in \cite{COletter04}. The accuracy in the temperature and
  column density estimates is limited by systematic instead of
  statistical errors (see text for a fuller
  discussion).\label{tab_disk_model}}
\begin{tabular}{llllllll}
\hline\hline
\noalign{\smallskip}
\multicolumn{1}{c}{Model}&\multicolumn{1}{c}{Species} & \multicolumn{1}{c}{$i$} & \multicolumn{1}{c}{$dv$} &   \multicolumn{1}{c}{$T_{\mathrm{gas}}$} & \multicolumn{1}{c}{$N$} & \multicolumn{1}{c}{$R_{\mathrm{min}}$} & \multicolumn{1}{c}{$R_{\mathrm{max}}$}\\
                       &  & \multicolumn{1}{c}{($\degr$)} & \multicolumn{1}{c}{(km s$^{-1}$)} & \multicolumn{1}{c}{(K)} & \multicolumn{1}{c}{(cm$^{-2}$)}&\multicolumn{1}{c}{(AU)}&\multicolumn{1}{c}{(AU)}\\
\hline
\noalign{\smallskip}
Disk & CO     & 27    & \phantom{2}5 & 1662 & 3.9 $\times$ 10$^{21}$ & 0.2       &  3.6\\
Slab & H$_2$O & n. a. & 25  & 1580 $\pm$ 510 & (2.9 $\pm$ 0.4) $\times$ 10$^{18}$ & n. a.     & n. a.\\
Disk & H$_2$O & 27    & \phantom{2}5 & 1550 $\pm$ 510 & (2.1 $\pm$ 0.4) $\times$ 10$^{18}$ & 2.0 $\pm$ 1 & 4.0 $\pm$ 1 \\
\noalign{\smallskip}
\hline
 \end{tabular}  

\end{center}
\end{table*}

% ------------------------------------------------------------------

\subsubsection{Disk model}

The second model includes broadening effects caused by a disk viewed
with an angle of 27$\degr$, as found by fitting the CO bandhead. The
slab model spectrum is convolved with a profile which takes into
account the Doppler shifts caused by hot water located in a Keplerian
disk. The disk extends from $R_{\mathrm{min}}$ till $R_{\mathrm{max}}$
which are two parameters of the model. The inclination is assumed
equal to the value obtained by fitting the first CO bandhead ($i$=27
$\degr$). Line emission from a Keplerian disk has a double-peak
profile.  The simple convolution procedure is justified for H$_2$O
column densities in the range 10$^{18}$--10$^{20}$ cm$^{-2}$ where the
optical depth is $\tau <<$1 (see Fig.~\ref{Fig_H2Otau}). The central
star is assumed to have mass of 6 $M_{\odot}$ \citep{Brgspec04}.

The dust sublimation radius, i.e. the radius where dust grains attain
1500~K, should provide a upper boundary to outer radius
$R_{\mathrm{max}}$ because beyond this radius dust opacity will
dominate over the gas opacity and water nor CO lines can be observed.
A total column density of gas of $\sim$10$^{25}$ cm$^{-2}$, derived by
multiplying the CO column density by 10$^4$ results in a dust opacity
of $\tau \sim$ 500 in the $K-$band using the extinction law of
\cite{Weingartner2001ApJ...548..296W}, which assumes a standard
gas-to-dust mass ratio, and $R_V$=3.1. A CO abundance of 10$^4$ is
reached for a large range of physical conditions (see
Section~\ref{chemical_models}). However, gas located in the upper
atmosphere of the disk can be warmer than the mid-plane.  The total
luminosity toward \object{08576nr292} is estimated to be around 10$^3$
$L_\odot$ for a star of mass $\simeq$ 6.0 $M_\odot$. A crude estimate
to the dust sublimation radius is $r_{\mathrm{d}}= 0.0346
\sqrt{L_*/L_\odot}$~AU $\simeq$~1~AU, where a dust sublimation
temperature of 1500~K is assumed.

The results of the fit by a disk model is displayed in
Fig.~\ref{Fig_disk_fit}.  The quality of the data and the resolution
do not permit an accurate constraint on the parameters such as
$R_{\mathrm{min}}$ and $R_{\mathrm{max}}$.  The gas temperature and
column density are however better bracketed because these two
parameters are set by the relative strength between lines at LTE. The
best fit parameters are provided in Table~\ref{tab_disk_model}
together with that derived from fitting the first CO bandhead.

\subsubsection{Discussion on the spectrum fitting}

The slab and disk models give similar values for the water gas
temperature and column density. This closeness can be ascribed to the
fact that the lines are optically thin. The intensity of optically
thin emission lines is not sensitive to the actual geometry of the
object. A mean value between the two models for the gas and column
density is adopted in this discussion. The resulting error on the gas
temperature is the scatter between the two temperatures and is much
smaller than in the formal error analysis (15~K instead of 510~K). The
water has a mean temperature of 1565 $\pm$ 15~K and a column density
of (2.5 $\pm$ 0.4) $\times$ 10$^{18}$ cm$^{-2}$.  Both models give
relatively similar values for the temperature and column density.
This proximity can be ascribed to the low optical thickness of the
water lines.

The fits using either a slab or a disk model match relatively well the
moderate signal-to-noise ratio spectrum except that both fail to
reproduce the strength of the peak at 2.287 $\mu$m (line 6) relative
to that of at 2.2875 (line 7). The upper level energy of both lines
are similar such that non-LTE effect can be excluded (see
Table~\ref{table_strongest_lines}).  Moreover the Einstein $A$
coefficient is slightly lower for line 6 than 7.  Therefore line 7
should appear stronger in most situations.  The most likely
explanation to this discrepancy is that the observed emission at 2.287
$\mu$m is a combination of hot water and H$_2$ 3--2 S(1) emission.
Indeed, a strong H$_2$ 1--0 S(1) line has been detected in
\object{08576nr292} at 2.218 $\mu$m \citep{Brgspec04}. The origin of
the H$_2$ lines is not well determined and a large contribution from
the extended H{\sc ii} region is probable. Alternatively, errors and
uncertainties in the Einstein-$A$ coefficients may exist and can be
sufficiently large to explain the bad fit around 2.287 $\mu$m.

The fit by a slab requires a higher turbulent width ($dv$=25 km
s$^{-1}$) than the disk model ($dv$=5 km s$^{-1}$). This difference
stems from the difficulty to constrain $dv$ with unresolved lines.
The unlikely high value of $dv$ for the slab model and the
simultaneous goodness of the fit to the CO and water lines with the
Keplerian disk model favor the latter interpretation for the geometry
of the hot CO and water emitting region. Synthetic spectra obtained
with a Keplerian disk and supersonic values for $dv$ give
unsatisfactory fits. Therefore we can rule out a turbulent Keplerian
disk.

  The water vapor column density of (2.5 $\pm$ 0.4) $\times$ 10$^{18}$
  cm$^{-2}$ and the gas temperature $T_{\mathrm{vapor}} \simeq$ 1565~K
  are close to the values found around the low-mass young star
  \object{SVS~13} by \cite{Carr2004ApJ...603..213C} which is a priori
  unexpected since \object{SVS~13} and \object{08576nr292} differ in
  mass. This similarity stems from that fact that the H$_2$O emission
  comes from a region very close to \object{SVS~13}.
  \cite{Carr2004ApJ...603..213C} find that the emitting area extents
  out to 0.3~AU while the water vapor is emitted from 2 to 4~AU in the
  disk around \object{08576nr292}. The hot gas is located at a smaller
  distance from the low-mass star than from the massive star.
  
  The water vapor temperature is about 100--200~K lower than that of
  the hot CO, in accordance with the findings of
  \cite{Carr2004ApJ...603..213C} for \object{SVS~13}, although the hot
  CO is much warmer around \object{SVS~13}. This lower temperature is
  also consistent with a large inner and outer radius
  (Table~\ref{tab_disk_model}).  Interestingly, the water temperature
  is close to the maximum dust sublimation temperature of 1500~K and
  the outer radius $R_{\mathrm{min}}$ is much larger than the dust
  sublimation radius of 1~AU, though the effect of large dust opacity
  in the observed spectrum is absent.  In principle the low contrast
  between the line and the continuum should have prevented any
  possible detection of the weak optically thin water emission lines
  in the presence of dust.  One possible explanation is the lack of
  dust grains close to the star at 1300-1500~K. The large luminosity
  of a B5 stars can push the dust grains well beyond the dust
  sublimation radius through radiation pressure.  Scaling the figures
  computed by \cite{Saija2003MNRAS.341.1239S} to the radius, mass and
  effective temperature of a B5 star, we obtain values for the ratios
  of radiation pressure force over gravity $\beta$ which are much
  larger than 10 for all types of grain compositions. We have not
  taken into account the viscous drag generated by the friction
  between the dust grains and the ambient gas which may be important
  in the inner dense disk. The absence of hot dust is also compatible
  with the relatively small near-infrared excess $E(J-K)$. As a result
  of the separation between the gas and the dust, the hot gas at
  $T\simeq$1500~K would be deprived of elemental oxygen which comes
  from the sublimation of silicate grains. A lower oxygen abundance in
  the gas phase may significant affects the chemistry. We postpone the
  discussion on the chemistry to Section~\ref{Oxygen_abundance}.
  Another possible explanation for the low continuum optical depth
  resides in the difference in dust grain lifetime. The maximum grain
  lifetime at a certain temperature is proportional to its radius.
  \citet{Lamy1974A&A....35..197L} estimated that the maximum lifetime
  of a 1 $\mu$m radius silicate grain at 1300~K is $\simeq$0.6 year
  and ten times shorter for a 0.1 $\mu$m radius grain. Therefore, we
  can envision that only the larger grains in the dust size
  distribution can survive in the region where $T$=1300-1500~K, which
  results in a lower dust opacity.  Alternatively, if the thermal
  coupling between the gas and the dust is inefficient, especially in
  the upper atmosphere of disks, the gas can be hotter than the dust
  \citep{Glassgold2004ApJ...615..972G}. In addition, the disk upper
  atmosphere can be relatively dust-free due to dust settling.  This
  latter possibility may be at work because the disk model with a
  subsonic turbulent velocity is favored.  Finally, it should be kept
  in mind that a combination of dust settling and size segregation is
  possible.

  We have demonstrated the presence of hot CO and H$_2$O in the inner
  4~AU of a Keplerian disk orbiting a B5 star. In the next section, a
  detailed chemical model is built in order to establish whether
  molecules can form and survive in the harsh environment around a hot
  star.

% ---------------------------------------------------------------------
\section{Chemical modeling}
\label{chemical_models}

\subsection{Pure gaseous disk}

The presence of relatively large amount of H$_2$O and CO in the inner
disk around young early type stars raises the question of their origin.

Early B stars are the main contributors to the mean interstellar
radiation field. A critical part of the UV spectrum is the
912-1000~\AA\ range used for the computation of H$_2$ and CO
photodissociation rates. We modeled the chemistry that occurs in the
HI regions using a pseudo-time dependent chemical code, i.e.  all
photons below 912~\AA\ have been absorbed. The main parameters of the
models are the gas temperature $T_{\mathrm{gas}}$ and density
$n_{\mathrm{H}}$, the strength of the UV field and the abundances of
the species at time $t$=0.  The temperature
($T_{\mathrm{gas}}$=1500--4000~K) and density range
($n_{\mathrm{H}}$=10$^{10}$--10$^{14}$ cm$^{-3}$) considered in our
modeling correspond to the values derived by
\cite{Lenorzer2004A&A...414..245L} when fitting the spectral energy
distribution of \object{NGC2024/IRS2}.

The radiation field experienced by a parcel of gas is dominated by the
emission from the central star diluted by the distance to the star and
by self-shielding. Because of the proximity to the central star, the
disk can be considered flat and the radiation flux received by the
disk surface at distance $R$ from the star of radius $R_*$ with
effective temperature $T_*$ is:
\begin{equation}
  F_{\mathrm{star}}=\frac{1}{2}\sigma T_*^4 \left(\frac{R_*}{R}\right)^2 \mu_0= \sigma T_*^4 W_{\mathrm{disk}},
\end{equation}
with
\begin{equation}
 \mu_0 = \frac{2}{3\pi}\left(\frac{R_*}{R}\right).
\end{equation}
$\mu_0$ is the cosine of the angle between the incoming stellar
radiation and the normal to the surface of the disk.  For a typical
B5V star, $R_*$= 3.9 $R_{\odot}$, therefore the dilution factor
$W_{\mathrm{disk}}$ at 2 AU is $W_{\mathrm{disk}}\sim$ 7.5 $\times$
10$^{-7}$ compared to $W_{\mathrm{ISM}}$=10$^{-14}$ for the mean
interstellar UV field.

% -------------------------------------------------------------------
   \begin{figure*}[ht]
     \centering
     \resizebox{\hsize}{!}{\includegraphics[angle=90]{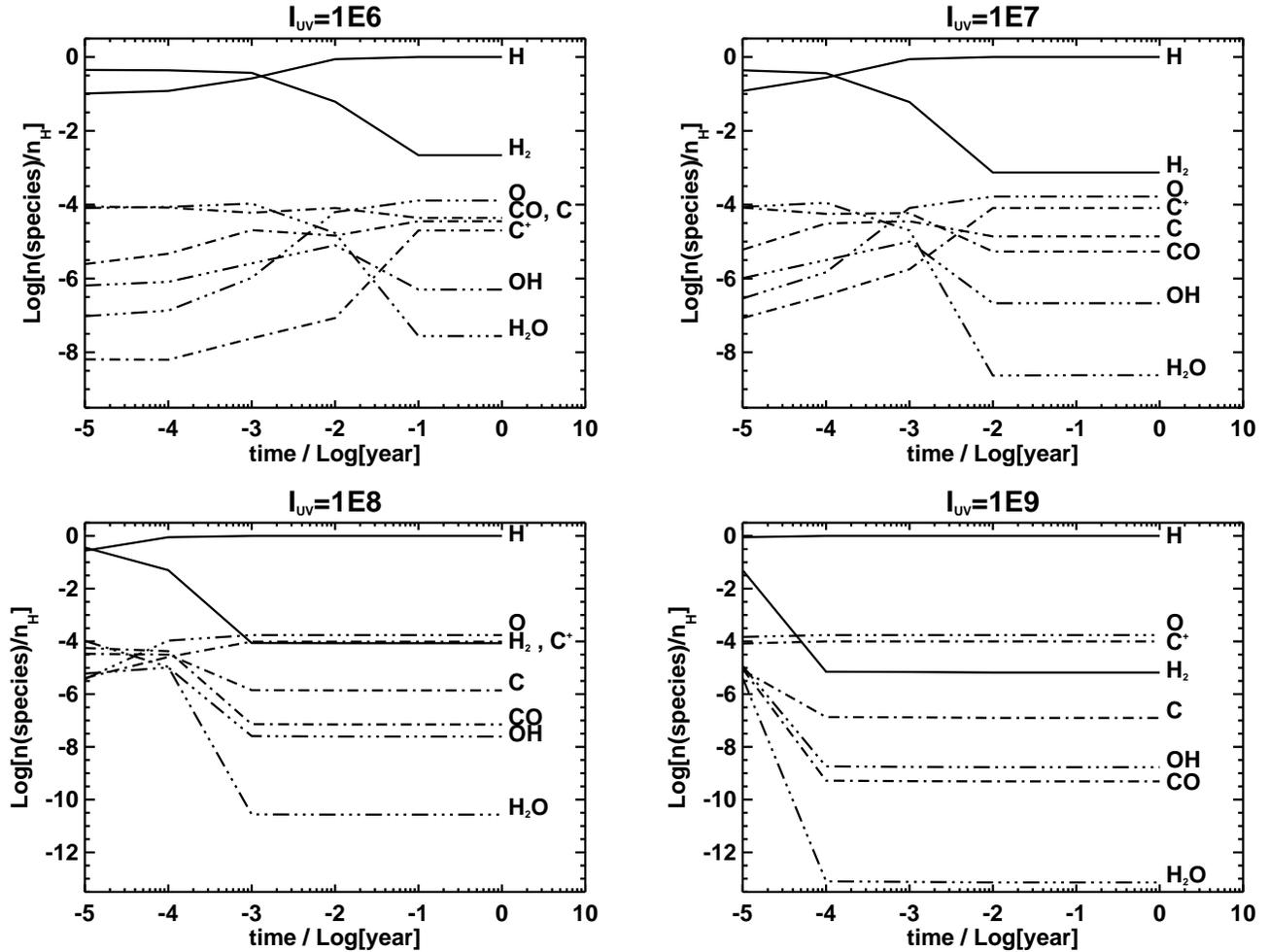}}
  \caption{Relative abundance of various species as a function of time and enhanced interstellar UV field $I_{\mathrm{UV}}$. The gas is free of dust grains, at temperature of 1600~K and has density of 10$^{12}$ cm$^{-3}$ as deduced from the fitting to the CO and H$_2$O emission lines. CO self-shielding was not included in these models.
\label{Fig_chem4plots}}
\end{figure*}
% -----------------------------------------------------------------

The photodissociation rates in the UMIST database are multiplied by a
factor $I_{\mathrm{UV}}=w_{\mathrm{disk}}/w_{\mathrm{ISM}}$=7.5
$\times$ 10$^{7}$.  The spectrum of the mean interstellar UV field is
close to that of a B0V star and it is assumed here that a unique
enhancement factor suffices to simulate the UV photodissociation. We
explore the effect of an enhanced UV field by factor
$I_{\mathrm{UV}}$=10$^6$--10$^9$ because of the uncertainty in the
actual spectral type of \object{08576nr292} \citep{Brgspec04} and in
the inner and outer radius of the emitting area. In photodominated
chemistry, an important parameter is the ratio
$I_{\mathrm{UV}}/n_{\mathrm{H}}$=10$^{-7}$--10$^{-2}$ which determines
the behavior of the chemistry. The energy balance of high-density
photodissociation regions is discussed by
\cite{Burton1990ApJ...365..620B}. The main cooling agents of gas at a
few thousand Kelvin are H$_2$ ro-vibrational lines, followed by H$_2$O
lines. The heating is primarily provided by H$_2$ vibrational heating.
Far UV photons will pump H$_2$ molecules to a bound excited electronic
state (Lyman and Werner bands), from which they will fluoresce down to
a high vibrational level of the ground electronic state. Most of the
time (85\%-90\%) the excited H$_2$ molecule will fluoresce back to an
low vibrational level while 10--15\% of the time the molecule will
dissociate. At the high densities considered in this study, the
collision of excited H$_2$ with other species contribute to the
heating of the gas and thermalization of the ro-vibrational levels. 

\subsection{Chemical model}

We select molecular species composed of five atoms or less and
containing the elements H, C, O, N, Si, Mg, Fe. The chemical reactions
are drawn from the compilation of UMIST rate99
\citep{LeTeuff2000A&AS..146..157L}. Contrary to the interstellar
chemistry in cold environments, we expect that reactions with
activation barrier, endothermic and three-body reactions dominate the
chemistry in inner hot disks. Therefore, besides the ion-molecules
reactions, two new types of reactions are taken into account: {\em
  termolecular} reactions, which are catalyzed bimolecular reactions
(i.e., 3-body reactions) and {\em collider} reactions, which are
collision-induced dissociations. To emphasize the likely importance of
3-body reactions, we provide below the dissociation sequence of water
by a third body which is believed to be the initiation reaction in the
so-called Bradford sequence.
\begin{equation}
\mathrm{H_2O + M \rightarrow OH + H}
\end{equation}
\begin{equation}
\mathrm{OH + CO \leftrightarrow CO_2 + H}
\end{equation}
\begin{equation}
\mathrm{H + H_2O \leftrightarrow OH + H_2}
\end{equation}
\begin{equation}
\mathrm{H + OH + M \rightarrow H_2O + M}
\end{equation}

\noindent where M designates a third body (H, He, H$_2$, ...). All together, 
the chemical network consists of 3299 reactions involving 301 species.
Most species have negligible abundance ($<$10$^{-30}$ w.r.t H). No
deuterated species are included in our study.

Another major difference with previous studies of chemistry in disks
around low- and intermediate-mass young stars
(\citealt{Aikawa1999ApJ...519..705A};
\citealt{Markwick2002A&A...385..632M};
\citealt{Nguyen2002A&A...387.1083N}), which focus on the chemistry
from a few AU ($R>$10~AU), is that the region modeled here is
dust-free. The first consequence is the absence of the H$_2$ formation
reaction on grain surfaces. The second is that far ultraviolet photons
are not absorbed by dust so that photodissociation is very effective.

% ------------------------------------------------------------
\begin{table}[!ht]
\centering
\caption{Initial abundances. The total oxygen budget is allowed to vary to explore the effect of different C/O elemental ratios. Water is assumed to be absent at $t$=0. \label{table_initial_abundances}}
\begin{tabular}{ll}
\hline
\hline
\noalign{\smallskip}
\multicolumn{1}{c}{species} & \multicolumn{1}{c}{n(species)/n(H)}\\
\hline
 \noalign{\smallskip}
 H       & 1.0 \\
 He      & 7.1 $\times$ 10$^{-2}$\\
 C       & 1.0 $\times$ 10$^{-4}$\\
 N       & 2.0 $\times$ 10$^{-5}$\\
 O       & 1.7 $\times$ 10$^{-4}$\\
 C$^{+}$ & 4.0 $\times$ 10$^{-6}$\\
 CO      & 8.5 $\times$ 10$^{-5}$\\
 H$_2$O  & 1.0 $\times$ 10$^{-30}$\\
\noalign{\smallskip}
\hline   
\end{tabular}
\end{table}
% ------------------------------------------------------------

\subsection{Numerical model}

The model is static and isothermal. The model solves numerically a
system of differential equations that simulate the formation and
destruction of atomic and molecular species. A few solvers of systems
of stiff first order differential equations were used and their
relative speed was checked.  The solvers are GEAR, VODE \citep{VODE}
and two alternative algorithms described by
\cite{Press1992nrfa.book.....P}.  All solvers give similar answers for
a few test models, ensuring that the results presented here are not
numerical noise caused by a specific solve. The adopted solver for the
results presented in this paper is VODE, the latest incarnation of the
backward-differentiation method popularized by GEAR.  This solver
gives the fastest results on most of the models. We cover a parameters
space in temperature (1500-5000~K), density ($n_\mathrm{
  H}$=10$^{10}$--10$^{14}$ cm$^{-3}$), UV field and CO self-shielding
factor. The initial abundance for the models is given in
table~\ref{table_initial_abundances}.  The cosmic-ray flux is set to
the interstellar value. Models were run up to one year which is much
than sufficient for the steady-state to be reached. (see next
section).

% ------------------------------------------------------------------
\section{Results of the chemical model calculations and discussion}
\label{results}

After presenting the results of the chemical model calculations, we
will describe the formation route of the main molecular species
(H$_2$, CO and water) in a hot gas immersed in a strong UV field. The
possibility that the molecules are formed in the outer part of the
disk and transported to the inner few AU is discussed. Then we
consider the effect of CO self-shielding against UV photodissociation
and non-standard C/O abundance ratio on the molecular abundances.
Finally the model predictions are compared to the observed abundances.

\subsection{Results}

\begin{figure*}[ht]
     \centering
     \resizebox{\hsize}{!}{\includegraphics[angle=90]{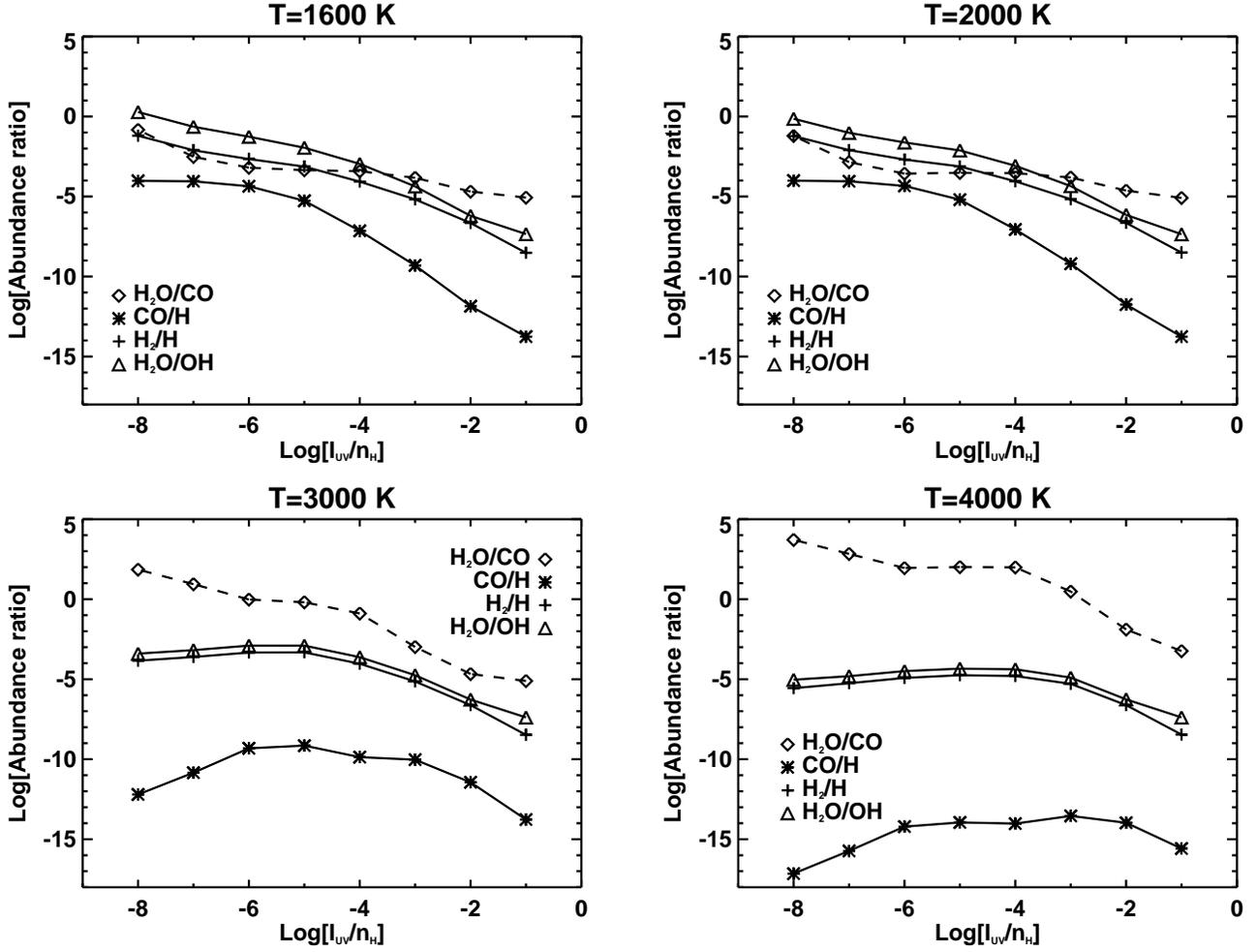}}
  \caption{Relative abundance of various species as a function of $I_{\mathrm{UV}}/n_{\mathrm{H}}$ for gas at temperature
    1600~K, 2000~K, 3000~K, 4000~K.
\label{Fig_ratios4plots}}
\end{figure*}

The abundance of H, H$_2$, O, OH, H$_2$O, C, C$^+$, CO, and N are
shown in Fig.~\ref{Fig_chem4plots} for a gas temperature of 1600~K,
gas density of 10$^{12}$ cm$^{-3}$ and four values of
$I_{\mathrm{UV}}$ (10$^6$, 10$^7$, 10$^8$, 10$^9$) and no CO nor H$_2$
self-shielding. First of all, it is clear that steady-state is reached
in less than a year.  This rapidity is reassuring because the matter
located in the inner few AU of disks rapidly falls onto the central
star. As a comparison, the rotational period of the gas at 1~AU around
a 6 $M_{\odot}$ star is $\simeq$ 150 days. In most cases, steady-state
is attained in a few days. The dominant species are the atomic H, O,
C, C$^+$, N.  Molecular hydrogen H$_2$, CO, OH and H$_2$O are the most
abundant radicals/molecules. Important abundance ratios such as
n(H)/n(H$_2$), n(OH)/n(H$_2$O) and n(CO)/n(H$_2$O) are displayed in
Fig.~\ref{Fig_ratios4plots} for different combinations of
$I_{\mathrm{UV}}/n_{\mathrm{H}}$ and $T_{\mathrm{gas}}$ (1600, 2000,
3000 and 4000~K). At $I_{\mathrm{UV}}/n_{\mathrm{H}}<10^{-5}$, the
chemistry is photo-regulated similar to a photodissociation region and
all newly formed molecules are instantaneously destroyed. In the
thermal region ($I_{\mathrm{UV}}/n_{\mathrm{H}}>10^{-5}$), the
abundance of CO, OH and H$_2$O reduces with increasing gas temperature
because, once formed, those molecules react further to synthesize
other species.  Therefore, the optimal temperature for molecular
abundance is around 2000$\pm$500~K.

% -----------------------------------------------------------------
\subsection{Hot dense gas chemical network}

\begin{figure}[ht]
     \centering
     \resizebox{\hsize}{!}{\includegraphics[]{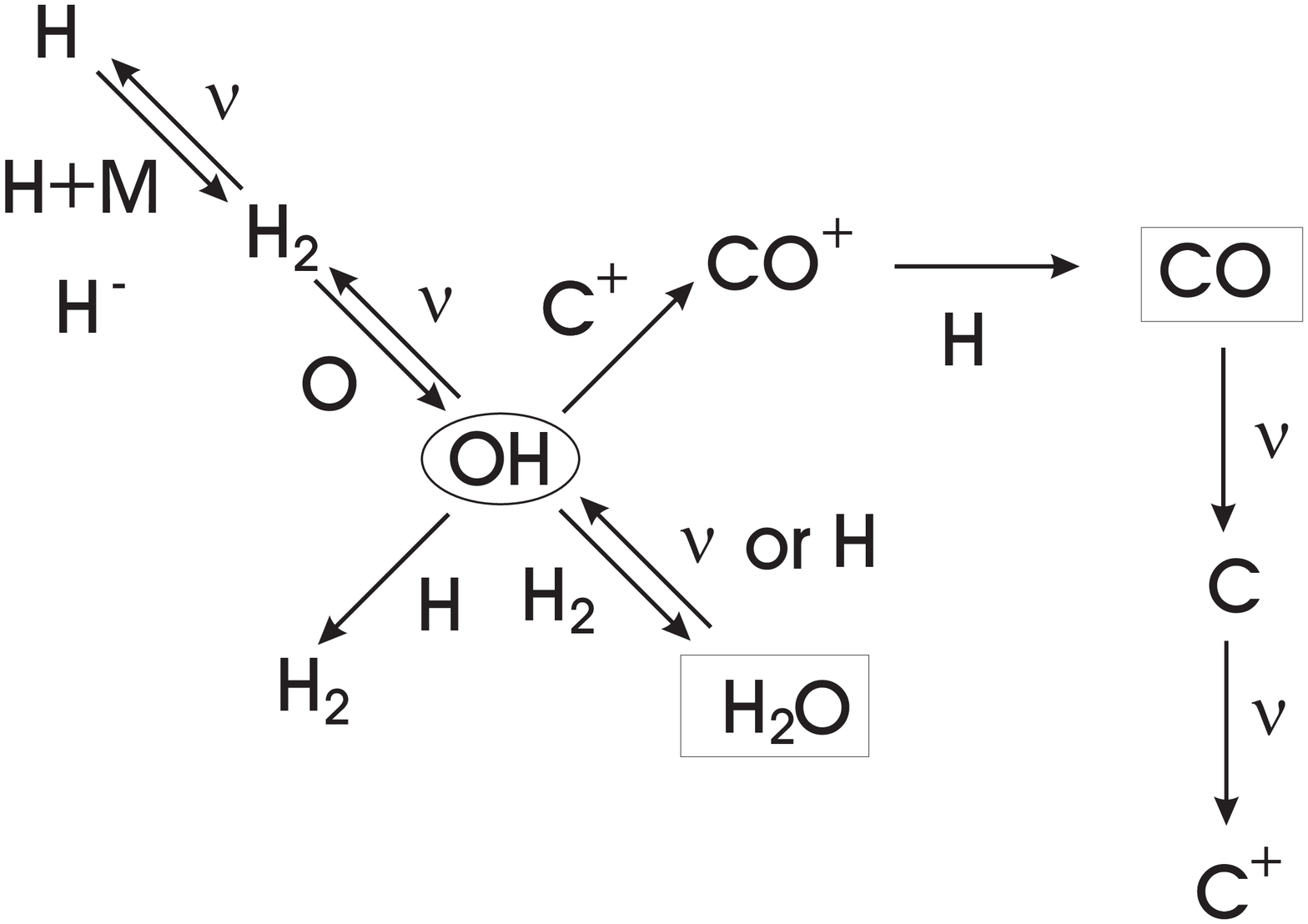}}
  \caption{Schematic representation of the proposed chemical network. OH radicals play
    a central role in this network. $\nu$ means photodissociation by
    ultraviolet photons and M a third body (M=H, He). The main routes
    leading to CO and H$_2$O are shown here.
\label{Fig_chem_network}}
\end{figure}

The chemical code monitors the main formation and destruction
reactions at designated times. The formation routes for the main
molecules are discussed below and a schematic is shown in
Fig.~\ref{Fig_chem_network}.  The chemical routes are different for
dense and extremely dense gases.  Below densities of around 10$^9$
cm$^{-3}$ and at high temperature, the chemistry is dominated by
two-body reactions with activation energy of the order of few thousand
Kelvin, whereas at very high densities, three-body reactions dominate.

% -------------------------------------------------------------------
% -----------------------------------------------------------------

% -------------------------------------------------------------------
% -----------------------------------------------------------------

\subsubsection{Formation of H$_2$}

Molecular hydrogen is the starting molecule for the chemical network.
In the absence of dust grains, H$_2$ is slowly synthesized in the gas
phase via three possible route depending on the density and
temperature. At moderate densities, the reaction
\begin{equation}
\mathrm{H^- +  H \rightarrow  H_2 + e-} 
\end{equation} 
\noindent dominates. This reaction has a flat rate of 1.30 $\times$ 10$^{-9}$ cm$^3$ s$^{-1}$.
At high temperature, H$_2$ formation is initiated by the radiative
association:
\begin{equation}
\mathrm{H^+ + H \rightarrow H_2^+ + photon} 
\end{equation}
\noindent which has an activation energy of 8500~K, followed by
\begin{equation}
\mathrm{H_2^+ + H \rightarrow H_2 + H^+}
\end{equation}
\noindent H$_2$ molecules further react with atomic oxygen or carbon to form OH or CH. At temperature higher than 3000~K, H$_2$ is dissociated by the collisions with
hydrogen atoms. This latter reaction has an activation energy of
55,000~K. At high densities ($n_{\mathrm{H}}>$10$^{12}$ cm$^{-3}$),
H$_2$ can also be formed by the three-body reaction:
\begin{equation}
\mathrm{H + H + H \rightarrow H_2 + H}
\end{equation}
\noindent which is a highly exothermic reaction, but the third hydrogen atom
carries away the excess energy. H$_2$ reacts swiftly with atomic
oxygen to from OH which is the key radical for the synthesis of CO and
H$_2$.

\subsubsection{CO formation and destruction}

CO molecules are readily formed for all values of the UV field
enhancement when the density reaches a value of 10$^{6}$ cm$^{-3}$ and
temperatures below 2500~K. The formation of CO is initiated by the
fast reaction at high temperature (activation energy of $E_\mathrm{
  a}$=3150~K)
\begin{equation}
\mathrm{H_2 + O \rightarrow OH + H}
\end{equation}
followed by the reaction:
\begin{equation}
\mathrm{C + OH \rightarrow CO + H}
\end{equation}
\noindent which has a rate of  1.10$\times$ 10$^{-10}$(T/100)$^{1/2}$ s$^{-1}$. The other channel
\begin{equation}
\mathrm{C + OH \rightarrow O + CH}
\end{equation}
has an activation energy of 14800~K.

Above 3000~K, destruction of CO is enhanced because reactions with
high barrier are activated. Especially, CO is chemically dissociated
at gas temperature greater than 3000~K by the endothermic reaction:
\begin{equation}
\mathrm{H + CO \rightarrow C + OH}
\end{equation}
This reaction is particularly important since hydrogen is mostly in
the atomic form.

At lower UV intensities, CO react with the OH$^+$ to form HCO$^+$.
Here OH$^+$ come from the reaction between H$_3^+$ and atomic oxygen.

At extreme densities, CO is formed directly from the neutral-neutral
addition C+O and the excess energy is carried away by a third body,
most likely a hydrogen atom. The main destruction reaction beside
photodissociation is the hydrogenation of CO leading to HCO. If the
abundance of H$_2$ is not sufficiently high
\begin{equation}
\mathrm{CO + H +M \rightarrow HCO + M}
\end{equation}
\noindent other possible reactions are
\begin{equation}
\mathrm{CO + H_2 + M \rightarrow HCO + H}
\end{equation}
\noindent where M is a third body (likely H or He). Alternatively, with low H$_2$ abundance, CO preferably react with OH to form CO$_2$:
\begin{equation}
\mathrm{CO + OH \rightarrow CO_2 + H}
\end{equation}
\noindent finally, the reaction:
\begin{equation}
\mathrm{HCO + H + M \rightarrow H_2CO + M}
\end{equation}
convert HCO into H$_2$CO. Therefore, HCO and H$_2$CO have
non-negligible abundance.

\subsubsection{Water molecules}

The chemical reactions leading to water molecules in cold and hot
regions are different. In quiescent molecular cloud, the gas
temperature is below 230~K and gas-phase chemistry gives typical
H$_2$O abundances of a few $\times$ 10$^{-7}$ (e.g.,
\citealt{Lee1996A&AS..119..111L};
\citealt{LeTeuff2000A&AS..146..157L}). Water is formed by a sequence
of ion-molecules reactions starting with O + H$_3^+$ or O$^+$ + H$_2$
leading to OH$^+$. Then rapid H-abstraction reactions with H$_2$
produce H$_3$O$^+$ which dissociatively recombines (with electrons) to
H$_2$O. Another potential source of gas phase water molecules is the
evaporation of icy grains in regions where $T>$100~K. The reaction
scheme in cold regions has been proven to be too effective in
producing H$_2$O when compared to observations of high-mass star
forming regions by \cite{Snell2000ApJ...539L..97S}. At high
temperature ($T_{\mathrm{gas}}>$230~K) and densities lower than
10$^{15}$ cm$^{-3}$, water molecules are predominately formed via the
radical-molecule reaction:
  \begin{equation}
    \mathrm{H_2 + OH \rightarrow H_2O + H}
  \end{equation}
  since OH is very abundant. The importance of this reaction has
  already been realized for water formation in shocks (e.g.,
  \citealt{Elitzur1979ApJ...229..560E};
  \citealt{Bergin1998ApJ...499..777B}). Water molecules are mostly
  destroyed by photodissociation and by reaction with C$^+$:
  \begin{equation}
    \mathrm{H_2O + C^+ \rightarrow HOC^+/HCO^+ + H}
  \end{equation}
  
  The Bradford sequence for the synthesis of water still plays a small
  role compared to the radical-molecule formation route described
  above at the densities attained in inner disks. The reaction network
  (CO and H$_2$O) discussed here and summarized in
  Fig.~\ref{Fig_chem_network} is similar to those found for
  post-$J$-shock chemistry \citep{Hollenbach1989ApJ...342..306H}.  It
  may therefore be difficult to distinguish between a post-shock or a
  quiescent high-temperature chemistry origin for many molecules.  The
  production of water vapor by sublimation of icy mantle when the dust
  temperature exceeds ~100~K is often advocated to explain large
  quantity of H$_2$O in the gas phase.

% -------------------------------------------------------------------
\begin{figure*}[ht]
   \centering
   \resizebox{\hsize}{!}{\includegraphics[angle=90]{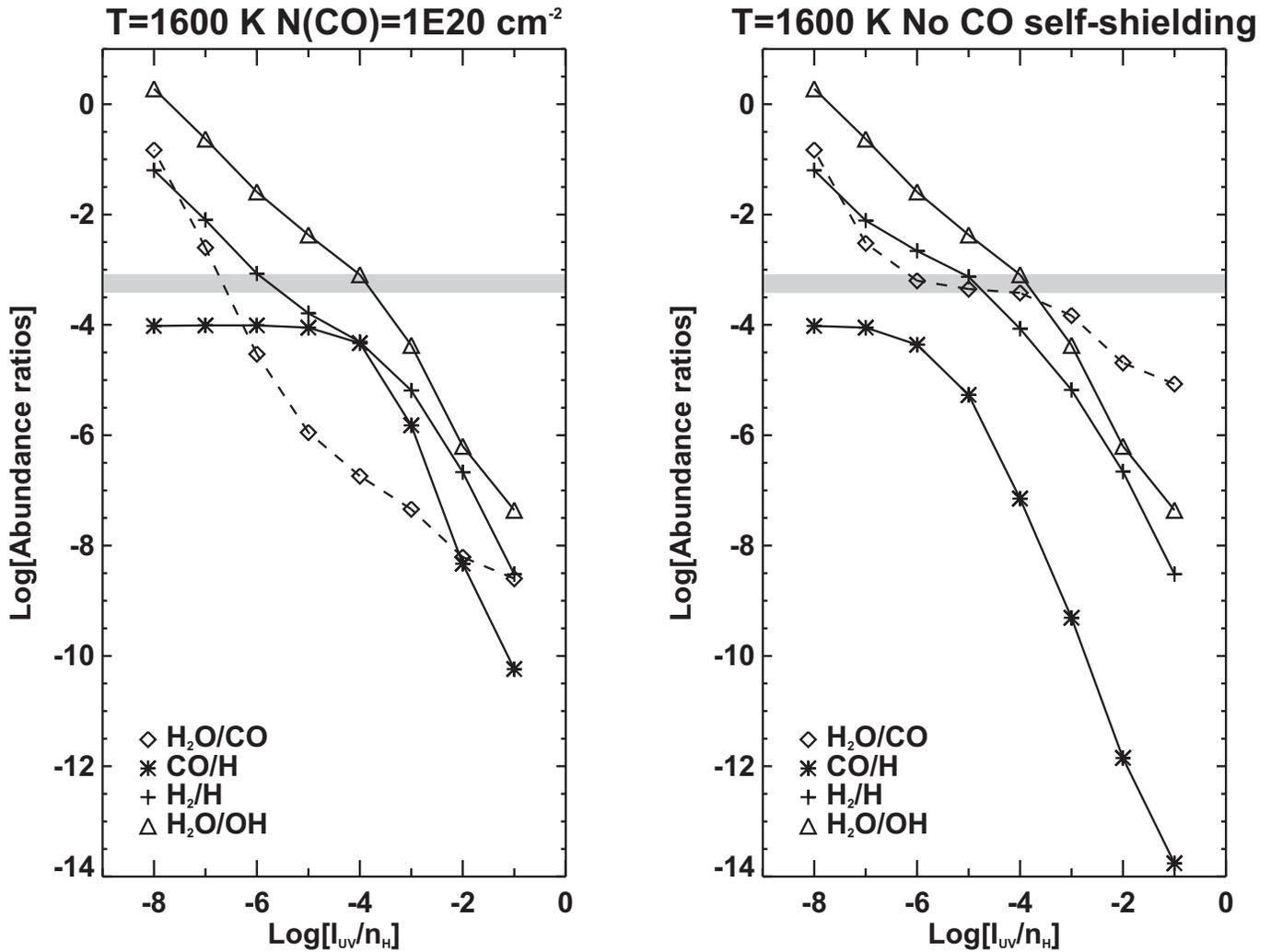}}
   \caption{Effect of CO self-shielding on abundance ratios. The left panel shows the addition of a strong self-shielding on the chemistry while the right panel is the same than the upper right panel in Fig.~\ref{Fig_ratios4plots}. As expected the changes happen only on the H$_2$O/CO and CO/H ratios. The observed value of H$_2$O/CO ($\sim$ 5.4 10$^{-4}$) is indicated by a grey strip.\label{fig_coshielding_ratios}}
 \end{figure*}

% ------------------------------------------------------------------

% -------------------------------------------------------------------
   \begin{figure}[h]
     \centering
     \resizebox{\hsize}{!}{\includegraphics[angle=90]{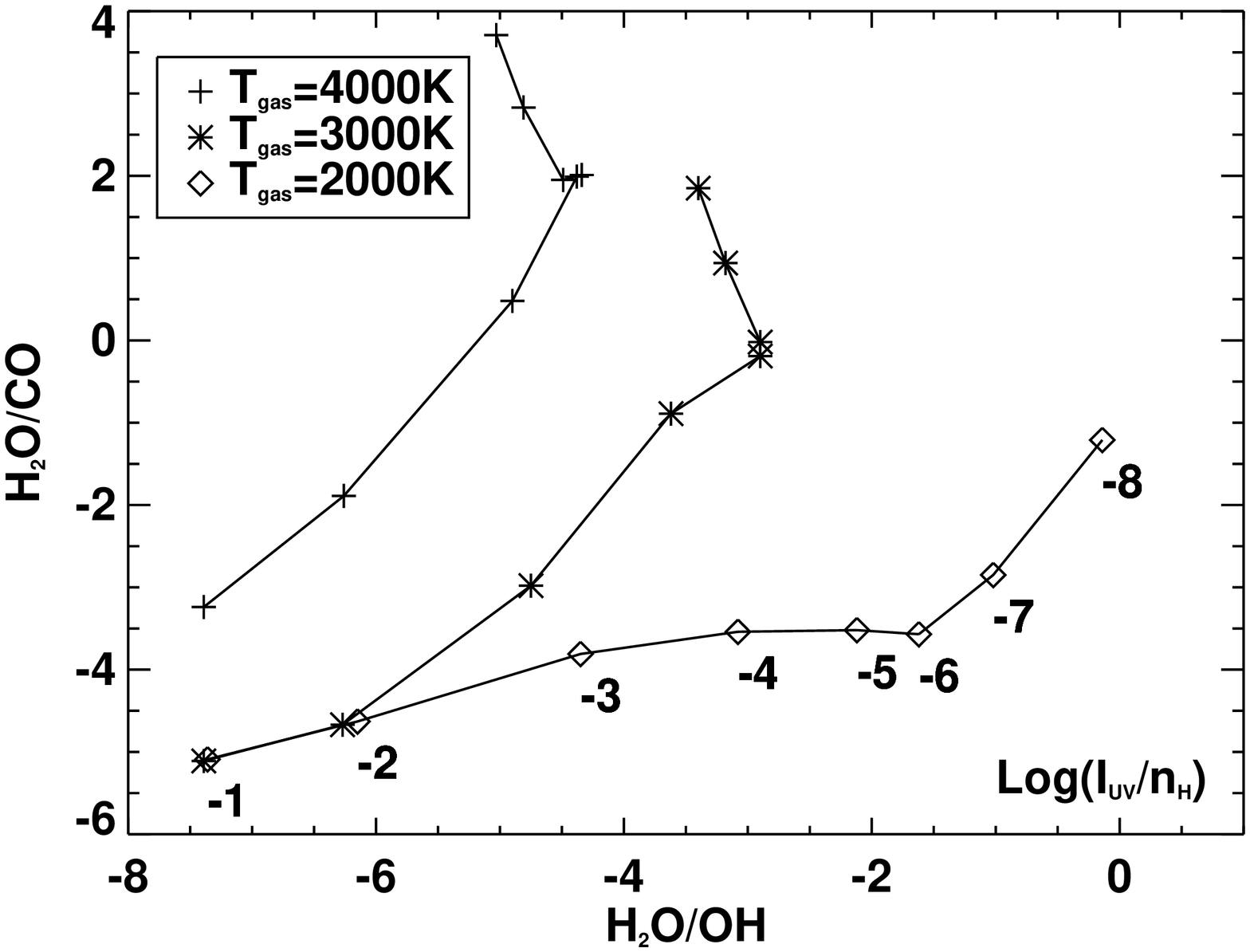}}
  \caption{H$_2$O/CO versus H$_2$O/OH diagram for various $I_{\mathrm{UV}}/n_{\mathrm{H}}$ and gas temperature. \label{Fig_ratio_vs_ratio}}
\end{figure}
% -----------------------------------------------------------------

\subsection{Endogenous versus exogenous formation of molecules}

Two main scenarii can explain the presence of molecules in the inner
disks around young massive stars.  The first model proposes that the
molecules are formed at large distances from the central object and
subsequently brought to its vicinity by accretion. We call this
possibility the exogenous scenario. In the exogenous model, molecular
species can also originate from the evaporation of icy grain mantle.
The second model states that molecules can be readily synthesized
locally in the inner disks. This is the endogenous model. All models
reach steady-state in at most one year and the effects of the initial
molecular abundances are minimal.

% -------------------------------------------------------------------
   \begin{figure}[th]
     \centering
     \resizebox{\hsize}{!}{\includegraphics[angle=90]{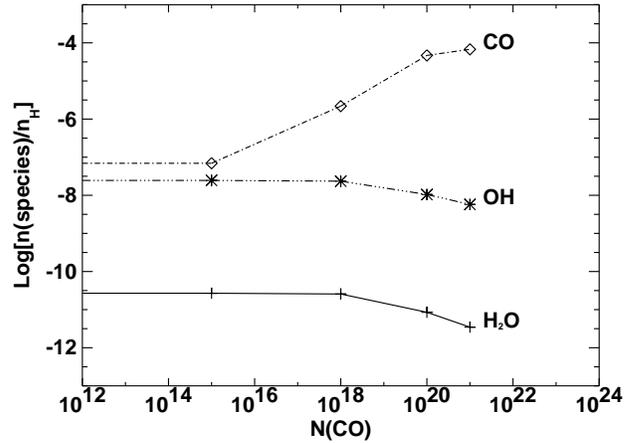}}
     \caption{Effect of the CO self-shielding on the relative abundances. The self-shielding factor
       depends on the CO column density $N$(CO). The parameters of the
       model are $T_{\mathrm{gas}}$=1600~K, $n_{\mathrm{H}}$=10$^{12}$
       cm$^{-3}$ and $I_{\mathrm{UV}}$=10$^{8}$.
       \label{fig_coshielding_effect}}
 \end{figure}
% -------------------------------------------------------------------
   \begin{figure}[ht]
     \centering
     \resizebox{\hsize}{!}{\includegraphics[angle=90]{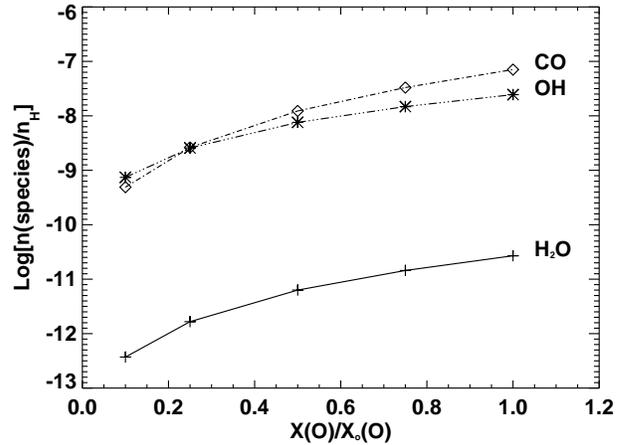}}
     \caption{Effect of the oxygen elemental abundance on the relative abundance of CO, OH and H$2$O. X$_{\mathrm{o}}$(O) is the nominal value of the Oxygen elemental abundance. The parameters of the model are $T_{\mathrm{gas}}$=1600~K, $n_{\mathrm{H}}$=10$^{12}$ cm$^{-3}$ and $I_{\mathrm{UV}}$=10$^{8}$. \label{fig_oxygen_effect}}
 \end{figure}
% -----------------------------------------------------------------

% ------------------------------------------------------------------
\subsection{Effect of CO self-shielding} \label{COshielding}

At large column densities, CO molecules can self-shield from
photodissociation (e.g., \citealt{vanDishoeck1988ApJ...334..771V}).
The nominal column of CO is 10$^{15}$ cm$^{-2}$ for a turbulent width
of 5 km s$^{-1}$ but this value was derived for gas at low temperature
and may not be valid for gas above 1500~K, because highly vibrational
excited absorption levels were not taken into account in models of CO
self-shielding. The nominal column of CO for self-shielding varies as
1/$\Delta v$. For a very turbulent gas with $\Delta v$= 30 km
s$^{-1}$, the nominal column of CO is increased by a factor $\sim$6.
Nevertheless, we ran models with increasing CO column density $N$(CO)
to account for probable increase of the value of the nominal CO for
self-shielding needed to simulate the condition in inner disks. The
other parameters are $T_{\mathrm{gas}}$=1600~K,
$n_{\mathrm{H}}$=10$^{12}$ cm$^{-3}$ and $I_{\mathrm{UV}}$=10$^{8}$.
Figure~\ref{fig_coshielding_effect} shows that above some threshold
value ($N$(CO)$>$10$^{15}$ cm$^{-2}$), CO molecules are self-protected
against photodissociation and its abundance can reach the cold
quiescent molecular cloud value of $\sim$10$^{-4}$.  Surprisingly, OH
and H$_2$O abundances seem to decrease when CO abundance rises
steeply. The H$_2$O/CO ratio drops dramatically from few 10$^{-4}$ to
few 10$^{-7}$. From Fig.~\ref{fig_coshielding_ratios}, it is clear
that the effect of CO self-shielding is most noticeable for
$I_{\mathrm{UV}}/n_{\mathrm{H}}>10^{-5}$ (the photodominated regime).
The CO abundance reaches similar value than in the thermal region,
i.e.  where $I_{\mathrm{UV}}/n_{\mathrm{H}}<10^{-5}$.

% -----------------------------------------------------------------
\subsection{Effect of C/O abundance ratio}\label{Oxygen_abundance}

By pushing the dust grains beyond the dust sublimation radius, the
radiation pressure will create a separation between gas and dust. The
segregation of the gas and dust results in an increase of the
elemental C/O ratio because the silicate grains, which contain
10--20\% of the available oxygen, do not sublimate. The effect of an
increase in the elemental C/O ratio above the solar value of 0.4 would
be to lower the water and molecular oxygen abundance. In principle,
the elemental abundances can be deduced from the study of photospheric
lines of the star if the star is not too embedded.
\cite{Terzieva1998ApJ...501..207T} discussed the different effects on
the chemistry caused by variations in the C/O elemental abundance but
their study is limited to low temperature chemistry where elements can
be depleted onto grains. Here we allow the oxygen elemental abundance
to vary from 0.1 to 1.0 times the nominal value listed in
Table~\ref{table_initial_abundances}. The model results at
steady-state are shown in Fig.~\ref{fig_oxygen_effect}. As expected,
the relative abundance of CO, OH and H$_2$O diminish with decreasing
oxygen elemental abundance.  The rate of decrease is similar for the
three species. Therefore, the relative abundances H$_2$O/CO and
H$_2$O/OH are not significantly affected by changing the oxygen
abundance. The simultaneous decrease in CO and H$_2$O abundance with
lower oxygen elemental abundance can be ascribed to the formation
route of both molecules with are formed via the radial OH. This is in
contrast with the chemistry that occurs in quiescent molecular clouds
where H$_2$O is synthesized at a much lower rate than CO, which
results in a higher CO/H$_2$O ratio as the oxygen elemental abundance
decreases.

\subsection{Comparison with the observations}

A large range of gas temperature (1600-4800~K) has been inferred from
the modeling of CO bandhead emission toward four high-mass young stars
\citep{COletter04}. The abundance of CO starts to decrease at
temperature above 2500~K owing to collisional dissociation of CO and
to the reaction of CO with OH to form CO$_2$.

The observed H$_2$O/CO ratio in \object{08576nr292} is 5.4 $\times$
10$^{-4}$. If CO self-shielding is taken into account, the value for
$I_{\mathrm{UV}}/n_{\mathrm{H}}$ is not well determined because the
theoretical curve for H$_2$O/CO is relatively flat around few times
10$^{-4}$ (upper left panel in Fig.~\ref{Fig_ratios4plots}).  The CO
abundance is 10$^{-7}$--10$^{-4}$ and H$_2$O abundance is
10$^{-11}$--10$^{-8}$. Interestingly, when CO is allowed to
self-shield, we can better infer
$I_{\mathrm{UV}}/n_{\mathrm{H}}\sim$10$^{-7}$ for a gas at $T$=1600~K
from the observed value of H$_2$O/CO since the ''degeneracy'' in the
H$_2$O/CO curve in the right panel of
Fig.~\ref{fig_coshielding_ratios} is lifted when CO self-shielding is
taken into account in the modeling.

The absolute abundances are low in comparison to that in the envelope
of high-mass young stellar objects \citep{Boonman2003A&A...406..937B}.
From the CO abundance in the unshielded case, we infer a total
hydrogen column density of 3.9 $\times$ 10$^{25}$--10$^{28}$ cm$^{-2}$
or surface density $\Sigma_0=$ 6.5--65,000 g cm$^{-2}$. In the case CO
self-shielding effects are included in the modeling, CO/H reaches the
maximum relative abundance of 10$^{-4}$ and the total hydrogen column
density amounts to 3.9 $\times$ 10$^{25}$ cm$^{-2}$ or a surface
density of $\Sigma_0=$ 6.5 g cm$^{-2}$. We provide a large possible
range for the derived total hydrogen column density because current CO
self-shielding models may not be applicable to the situation
encountered in inner disks. However, it is noteworthy that the upper
range of column density ($N$(H)$>$10$^{26}$ cm$^{-2}$) would imply
large opacity in the near-infrared due to H$^-$, which would prevent
the detection of optically thin H$_2$O lines. The surface density is
comparable to the values adopted in disk models around T~Tauri and
Herbig~Ae stars which is $\Sigma_0 \sim$10$^{2}$-10$^{4}$ g cm$^{-2}$
at 1~AU.

We summarize the chemical modeling in Fig.~\ref{Fig_ratio_vs_ratio}.
This figure shows the {\bf H$_2$O/CO} versus H$_2$O/OH abundance
ratio.  The figure provides a simple way to deduce physical parameters
such as $T_{\mathrm{gas}}$ and $I_{\mathrm{UV}}/n_{\mathrm{H}}$ from
CO, OH and H$_2$O abundance. Unfortunately, OH lines were not covered
in the $K$-band spectrum of \object{08576nr292}.

\subsection{Comparison with the inner disk chemistry around low-mass young stars}

\cite{Glassgold2004ApJ...615..972G} studied the thermal-chemical
structure of the inner disks around young low mass stars. Although in
low-mass young stars X-rays drive the chemistry rather than the UV,
they found that the disk atmosphere may contain a significant amount
of warm ($T$=500--2000~K) CO gas similar to our results. This
similarity is not surprising since at the high temperatures and
densities encountered in inner disks around low- and high-mass young
stars, the chemistry is dominated by the same neutral-neutral
reactions. At very high densities, the UV and X-ray photons provide
the energy needed to maintain a high temperature.  The
photodissociation by X-ray or UV photons of the molecules are
compensated by rapid formation in dust-free regions. In a dust-free
region X-ray and UV photons are absorbed by gas while in a dust-rich
region UV radiation is rapidely attenuated by dust absorption while
the X-ray can penetrate deeper to influence the ionization fraction
(e.g., \citealt{Semenov2003A&A...410..611S}).

% --------------------------------------
\section{Conclusion and future prospects}
\label{conclusion}

We have shown that hot-water vapor emission is present in the close
vicinity ($R<$5~AU) of the high-mass young stellar object
\object{08576nr292}. The water vapor as well as the hot CO are likely
located in a disk in Keplerian rotation.

The detection of optically thin CO and H$_2$O lines at 1500-1600~K
suggests that dust grains are absent in that temperature range in the
disk. This may stem from the fact that dust grains are pushed further
out by radiation pressure or that most dust has evaporated.  The
region probed by hot CO and H$_2$O is limited to the inner few AU of
the disk. Alternatively, the gas and dust may be thermaly decoupled
with the gas being much warmer than the dust.

To probe cooler gas, which is located between 1 and $\sim$50 ~AU,
fundamental transitions of CO and lower $J$ lines of H$_2$O appear to
be more appropriate although the line over continuum contrast would be
smaller owing to the warm dust emission. Nevertheless, hot lines allow
the gas-rich and probably dust-poor region to be investigated. Higher
resolution spectrometers ($R$=100,000 or $\Delta v$=3 km s$^{-1}$)
such as {\em CRIRES} planned for the {\em VLT}, will resolve the
individual H$_2$O line profiles.  The combination of medium resolution
($R$=10,000) and high spatial resolution achieved by {\em AMBER} will
help to determine the exact location of both the CO and H$_2$O
emissions.

Chemical models have been developed for the dense and hot inner disk
around young high-mass stars. The chemistry is dominated by
neutral-neutral reactions and steady-state is rapidly attained.  H$_2$
formation is hampered by the lack of dust grains and reach relatively
low abundance. The radical OH is central to the chemical network
playing the same role than H$_3^+$ for cold chemistry. The molecular
abundances depend in general on the gas temperature and the UV flux
over density ratio and show a large range of values.
Photodissociation affects more H$_2$, OH and H$_2$O than CO because
the latter species can self-shield very efficiently.  The observed
H$_2$O/CO can be reproduced at the temperature of the gas (1600~K) and
$I_{\mathrm{UV}}/n_{\mathrm{H}}\sim$10$^{-6}$.  Observations of OH are
warranted to allow better constraints on the UV radiation field.
Future work on the modeling side includes the combination of the
chemical model presented here with a dynamic disk model around young
high-mass stars.

The detection of molecules such as CO and H$_2$O at high temperature
in the close vicinity of a young massive young stellar objects opens
the possibility to study the physics and chemistry in the region where
matter from the disk falls onto the star.

\begin{acknowledgements}
  WFT was supported by NWO grant 614.041.005 during his stay in
  Amsterdam.  The authors thank the VLT staff for performing the
  observations in Service mode. Comments and suggestions from Rens
  Waters and Lex Kaper are well appreciated.
\end{acknowledgements}

%%%%%%%%%%%%%%%%%%%%%%%%%%%%%%%%%%%%%%%%%%%%%%%%%%%%%%%%%%%%%%%%%

\bibliographystyle{aa}
\bibliography{vapor}  

\begin{thebibliography}{58}
\expandafter\ifx\csname natexlab\endcsname\relax\def\natexlab#1{#1}\fi

\bibitem[{{Aikawa} {et~al.}(1999){Aikawa}, {Umebayashi}, {Nakano}, \&
  {Miyama}}]{Aikawa1999ApJ...519..705A}
{Aikawa}, Y., {Umebayashi}, T., {Nakano}, T., \& {Miyama}, S.~M. 1999, \apj,
  519, 705

\bibitem[{{Allard} {et~al.}(2001){Allard}, {Hauschildt}, {Alexander},
  {Tamanai}, \& {Schweitzer}}]{Allard2001ApJ...556..357A}
{Allard}, F., {Hauschildt}, P.~H., {Alexander}, D.~R., {Tamanai}, A., \&
  {Schweitzer}, A. 2001, \apj, 556, 357

\bibitem[{{Bergin} {et~al.}(2000){Bergin}, {Melnick}, {Stauffer}, {Ashby},
  {Chin}, {Erickson}, {Goldsmith}, {Harwit}, {Howe}, {Kleiner}, {Koch},
  {Neufeld}, {Patten}, {Plume}, {Schieder}, {Snell}, {Tolls}, {Wang},
  {Winnewisser}, \& {Zhang}}]{Bergin2000ApJ...539L.129B}
{Bergin}, E.~A., {Melnick}, G.~J., {Stauffer}, J.~R., {et~al.} 2000, \apjl,
  539, L129

\bibitem[{{Bergin} {et~al.}(1998){Bergin}, {Neufeld}, \&
  {Melnick}}]{Bergin1998ApJ...499..777B}
{Bergin}, E.~A., {Neufeld}, D.~A., \& {Melnick}, G.~J. 1998, \apj, 499, 777

\bibitem[{{Bernath}(1995)}]{Bernath_book1995}
{Bernath}, P.~F. 1995, {Spectra of Atoms and Molecules} (Oxford University
  Press, N.Y., 400 p.)

\bibitem[{{Bevington} \& {Robinson}(2003)}]{Bevington2003drea.book.....B}
{Bevington}, P.~R. \& {Robinson}, D.~K. 2003, {Data reduction and error
  analysis for the physical sciences} (Data reduction and error analysis for
  the physical sciences, 3rd ed., by Philip R.~Bevington, and Keith
  D.~Robinson.~Boston, MA: McGraw-Hill, ISBN 0-07-247227-8, 2003.)

\bibitem[{{Bik} {et~al.}(2004){Bik}, {Kaper}, {Waters}, \&
  {Hanson}}]{Brgspec04}
{Bik}, A., {Kaper}, L., {Waters}, L.~B.~F.~M., \& {Hanson}, M.~M. 2004,
  submitted to \aap.

\bibitem[{{Bik} \& {Thi}(2004)}]{COletter04}
{Bik}, A. \& {Thi}, W.~F. 2004, \aap, 427, L13

\bibitem[{{Blake} \& {Boogert}(2004)}]{Blake2004ApJ...606L..73B}
{Blake}, G.~A. \& {Boogert}, A.~C.~A. 2004, \apjl, 606, L73

\bibitem[{{Blum} {et~al.}(2005){Blum}, {Barbosa}, {Damineli}, \&
  {Conti}}]{Blum2004}
{Blum}, R.~D., {Barbosa}, C.~L., {Damineli}, A., \& {Conti}, P.~I. 2005, \apj,
  astro-ph 0409190

\bibitem[{{Bonnell} \& {Bate}(2002)}]{Bonnell2002MNRAS.336..659B}
{Bonnell}, I.~A. \& {Bate}, M.~R. 2002, \mnras, 336, 659

\bibitem[{{Boonman} {et~al.}(2003){Boonman}, {Doty}, {van Dishoeck}, {Bergin},
  {Melnick}, {Wright}, \& {Stark}}]{Boonman2003A&A...406..937B}
{Boonman}, A.~M.~S., {Doty}, S.~D., {van Dishoeck}, E.~F., {et~al.} 2003, \aap,
  406, 937

\bibitem[{{Brittain} {et~al.}(2003){Brittain}, {Rettig}, {Simon}, {Kulesa},
  {DiSanti}, \& {Dello Russo}}]{Brittain2003ApJ...588..535B}
{Brittain}, S.~D., {Rettig}, T.~W., {Simon}, T., {et~al.} 2003, \apj, 588, 535

\bibitem[{{Brown} {et~al.}(1989){Brown}, {Byrne}, \& {Hindmarsh}}]{VODE}
{Brown}, P.~N., {Byrne}, G.~D., \& {Hindmarsh}, A.~C. 1989, SIAM J. Sci. Stat.
  Comput., 10, 1038

\bibitem[{{Burton} {et~al.}(1990){Burton}, {Hollenbach}, \&
  {Tielens}}]{Burton1990ApJ...365..620B}
{Burton}, M.~G., {Hollenbach}, D.~J., \& {Tielens}, A.~G.~G.~M. 1990, \apj,
  365, 620

\bibitem[{{Carr} {et~al.}(2004){Carr}, {Tokunaga}, \&
  {Najita}}]{Carr2004ApJ...603..213C}
{Carr}, J.~S., {Tokunaga}, A.~T., \& {Najita}, J. 2004, \apj, 603, 213

\bibitem[{{Chandler} {et~al.}(1995){Chandler}, {Carlstrom}, \&
  {Scoville}}]{Chandler1995ApJ...446..793C}
{Chandler}, C.~J., {Carlstrom}, J.~E., \& {Scoville}, N.~Z. 1995, \apj, 446,
  793

\bibitem[{{Charbonneau}(1995)}]{Charbonneau1995ApJS..101..309C}
{Charbonneau}, P. 1995, \apjs, 101, 309

\bibitem[{{Churchwell}(2002)}]{Churchwell2002ARA&A..40...27C}
{Churchwell}, E. 2002, \araa, 40, 27

\bibitem[{{Elitzur}(1979)}]{Elitzur1979ApJ...229..560E}
{Elitzur}, M. 1979, \apj, 229, 560

\bibitem[{{Glassgold} {et~al.}(2004){Glassgold}, {Najita}, \&
  {Igea}}]{Glassgold2004ApJ...615..972G}
{Glassgold}, A.~E., {Najita}, J., \& {Igea}, J. 2004, \apj, 615, 972

\bibitem[{{Helmich} {et~al.}(1996){Helmich}, {van Dishoeck}, {Black}, {de
  Graauw}, {Beintema}, {Heras}, {Lahuis}, {Morris}, \&
  {Valentijn}}]{Helmich1996A&A...315L.173H}
{Helmich}, F.~P., {van Dishoeck}, E.~F., {Black}, J.~H., {et~al.} 1996, \aap,
  315, L173

\bibitem[{{Hollenbach} \& {McKee}(1989)}]{Hollenbach1989ApJ...342..306H}
{Hollenbach}, D. \& {McKee}, C.~F. 1989, \apj, 342, 306

\bibitem[{{Jijina} \& {Adams}(1996)}]{Jijina1996ApJ...462..874J}
{Jijina}, J. \& {Adams}, F.~C. 1996, \apj, 462, 874

\bibitem[{{Kamp} {et~al.}(2003){Kamp}, {van Zadelhoff}, {van Dishoeck}, \&
  {Stark}}]{Kamp2003A&A...397.1129K}
{Kamp}, I., {van Zadelhoff}, G.-J., {van Dishoeck}, E.~F., \& {Stark}, R. 2003,
  \aap, 397, 1129

\bibitem[{{Kaper} {et~al.}(2005){Kaper}, {Bik}, {Comer\'on}, \&
  {Hanson}}]{Kaper05}
{Kaper}, L., {Bik}, A., {Comer\'on}, F., \& {Hanson}, M.~M. 2005, to be
  sumbitted to \aap.

\bibitem[{{Kaufman} \& {Neufeld}(1996)}]{Kaufman1996ApJ...456..611K}
{Kaufman}, M.~J. \& {Neufeld}, D.~A. 1996, \apj, 456, 611

\bibitem[{{Kraus} {et~al.}(2000){Kraus}, {Kr{\" u}gel}, {Thum}, \&
  {Geballe}}]{Kraus00}
{Kraus}, M., {Kr{\" u}gel}, E., {Thum}, C., \& {Geballe}, T.~R. 2000, \aap,
  362, 158

\bibitem[{{Lamy}(1974)}]{Lamy1974A&A....35..197L}
{Lamy}, P.~L. 1974, \aap, 35, 197

\bibitem[{{Le Teuff} {et~al.}(2000){Le Teuff}, {Millar}, \&
  {Markwick}}]{LeTeuff2000A&AS..146..157L}
{Le Teuff}, Y.~H., {Millar}, T.~J., \& {Markwick}, A.~J. 2000, \aaps, 146, 157

\bibitem[{{Lee} {et~al.}(1996){Lee}, {Bettens}, \&
  {Herbst}}]{Lee1996A&AS..119..111L}
{Lee}, H.-H., {Bettens}, R.~P.~A., \& {Herbst}, E. 1996, \aaps, 119, 111

\bibitem[{{Lenorzer} {et~al.}(2004){Lenorzer}, {Bik}, {de Koter}, {Kurtz},
  {Waters}, {Kaper}, {Jones}, \& {Geballe}}]{Lenorzer2004A&A...414..245L}
{Lenorzer}, A., {Bik}, A., {de Koter}, A., {et~al.} 2004, \aap, 414, 245

\bibitem[{{Lewis} \& {Prinn}(1980)}]{Lewis1980ApJ...238..357L}
{Lewis}, J.~S. \& {Prinn}, R.~G. 1980, \apj, 238, 357

\bibitem[{{Liseau} {et~al.}(1992){Liseau}, {Lorenzetti}, {Nisini}, {Spinoglio},
  \& {Moneti}}]{Liseau92}
{Liseau}, R., {Lorenzetti}, D., {Nisini}, B., {Spinoglio}, L., \& {Moneti}, A.
  1992, \aap, 265, 577

\bibitem[{{Markwick} {et~al.}(2002){Markwick}, {Ilgner}, {Millar}, \&
  {Henning}}]{Markwick2002A&A...385..632M}
{Markwick}, A.~J., {Ilgner}, M., {Millar}, T.~J., \& {Henning}, T. 2002, \aap,
  385, 632

\bibitem[{{Melnick} {et~al.}(2000){Melnick}, {Ashby}, {Plume}, {Bergin},
  {Neufeld}, {Chin}, {Erickson}, {Goldsmith}, {Harwit}, {Howe}, {Kleiner},
  {Koch}, {Patten}, {Schieder}, {Snell}, {Stauffer}, {Tolls}, {Wang},
  {Winnewisser}, \& {Zhang}}]{Melnick2000ApJ...539L..87M}
{Melnick}, G.~J., {Ashby}, M.~L.~N., {Plume}, R., {et~al.} 2000, \apjl, 539,
  L87

\bibitem[{{Mundy} {et~al.}(2000){Mundy}, {Looney}, \&
  {Welch}}]{Mundy2000prpl.conf..355M}
{Mundy}, L.~G., {Looney}, L.~W., \& {Welch}, W.~J. 2000, Protostars and Planets
  IV, 355

\bibitem[{{Najita} {et~al.}(1996){Najita}, {Carr}, {Glassgold}, {Shu}, \&
  {Tokunaga}}]{Najita1996ApJ...462..919N}
{Najita}, J., {Carr}, J.~S., {Glassgold}, A.~E., {Shu}, F.~H., \& {Tokunaga},
  A.~T. 1996, \apj, 462, 919

\bibitem[{{Najita} {et~al.}(2003){Najita}, {Carr}, \&
  {Mathieu}}]{Najita2003ApJ...589..931N}
{Najita}, J., {Carr}, J.~S., \& {Mathieu}, R.~D. 2003, \apj, 589, 931

\bibitem[{{Natta} {et~al.}(2000){Natta}, {Grinin}, \&
  {Mannings}}]{Natta2000prpl.conf..559N}
{Natta}, A., {Grinin}, V., \& {Mannings}, V. 2000, Protostars and Planets IV,
  559

\bibitem[{{Nguyen} {et~al.}(2002){Nguyen}, {Viti}, \&
  {Williams}}]{Nguyen2002A&A...387.1083N}
{Nguyen}, T.~K., {Viti}, S., \& {Williams}, D.~A. 2002, \aap, 387, 1083

\bibitem[{{Press} {et~al.}(1992){Press}, {Teukolsky}, {Vetterling}, \&
  {Flannery}}]{Press1992nrfa.book.....P}
{Press}, W.~H., {Teukolsky}, S.~A., {Vetterling}, W.~T., \& {Flannery}, B.~P.
  1992, {Numerical recipes in FORTRAN. The art of scientific computing}
  (Cambridge: University Press, |c1992, 2nd ed.)

\bibitem[{{Rothman} {et~al.}(2003){Rothman}, {Barbe}, {Benner}, {Brown},
  {Camy-Peyret}, {Carleer}, {Chance}, {Clerbaux}, {Dana}, {Devi}, {Fayt},
  {Flaud}, {Gamache}, {Goldman}, {Jacquemart}, {Jucks}, {Lafferty}, {Mandin},
  {Massie}, {Nemtchinov}, {Newnham}, {Perrin}, {Rinsland}, {Schroeder},
  {Smith}, {Smith}, {Tang}, {Toth}, {Vander Auwera}, {Varanasi}, \&
  {Yoshino}}]{Rothman2003JQSRT..82....5R}
{Rothman}, L.~S., {Barbe}, A., {Benner}, D.~C., {et~al.} 2003, Journal of
  Quantitative Spectroscopy and Radiative Transfer, 82, 5

\bibitem[{{Saija} {et~al.}(2003){Saija}, {Iat{\` i}}, {Giusto}, {Borghese},
  {Denti}, {Aiello}, \& {Cecchi-Pestellini}}]{Saija2003MNRAS.341.1239S}
{Saija}, R., {Iat{\` i}}, M.~A., {Giusto}, A., {et~al.} 2003, \mnras, 341, 1239

\bibitem[{{Scoville} {et~al.}(1983){Scoville}, {Kleinmann}, {Hall}, \&
  {Ridgway}}]{Scoville83}
{Scoville}, N., {Kleinmann}, S.~G., {Hall}, D.~N.~B., \& {Ridgway}, S.~T. 1983,
  \apj, 275, 201

\bibitem[{{Semenov} {et~al.}(2003){Semenov}, {Henning}, {Helling}, {Ilgner}, \&
  {Sedlmayr}}]{Semenov2003A&A...410..611S}
{Semenov}, D., {Henning}, T., {Helling}, C., {Ilgner}, M., \& {Sedlmayr}, E.
  2003, \aap, 410, 611

\bibitem[{{Semenov} {et~al.}(2004){Semenov}, {Wiebe}, \&
  {Henning}}]{Semenov2004A&A...417...93S}
{Semenov}, D., {Wiebe}, D., \& {Henning}, T. 2004, \aap, 417, 93

\bibitem[{{Snell} {et~al.}(2000){Snell}, {Howe}, {Ashby}, {Bergin}, {Chin},
  {Erickson}, {Goldsmith}, {Harwit}, {Kleiner}, {Koch}, {Neufeld}, {Patten},
  {Plume}, {Schieder}, {Stauffer}, {Tolls}, {Wang}, {Winnewisser}, {Zhang}, \&
  {Melnick}}]{Snell2000ApJ...539L..97S}
{Snell}, R.~L., {Howe}, J.~E., {Ashby}, M.~L.~N., {et~al.} 2000, \apjl, 539,
  L97

\bibitem[{{Tennyson} {et~al.}(2001){Tennyson}, {Zobov}, {Williamson},
  {Polyansky}, \& {Bernath}}]{Tennyson2001JPCRD..30..735T}
{Tennyson}, J., {Zobov}, N.~F., {Williamson}, R., {Polyansky}, O.~L., \&
  {Bernath}, P.~F. 2001, Journal of Physical and Chemical Reference Data, 30,
  735

\bibitem[{{Terzieva} \& {Herbst}(1998)}]{Terzieva1998ApJ...501..207T}
{Terzieva}, R. \& {Herbst}, E. 1998, \apj, 501, 207

\bibitem[{{van Dishoeck} \& {Black}(1988)}]{vanDishoeck1988ApJ...334..771V}
{van Dishoeck}, E.~F. \& {Black}, J.~H. 1988, \apj, 334, 771

\bibitem[{{Walsh} {et~al.}(1999){Walsh}, {Burton}, {Hyland}, \&
  {Robinson}}]{Walsh99}
{Walsh}, A.~J., {Burton}, M.~G., {Hyland}, A.~R., \& {Robinson}, G. 1999,
  MNRAS, 309, 905

\bibitem[{{Waters} \& {Waelkens}(1998)}]{Waters1998ARA&A..36..233W}
{Waters}, L.~B.~F.~M. \& {Waelkens}, C. 1998, \araa, 36, 233

\bibitem[{{Weingartner} \& {Draine}(2001)}]{Weingartner2001ApJ...548..296W}
{Weingartner}, J.~C. \& {Draine}, B.~T. 2001, \apj, 548, 296

\bibitem[{{Wolfire} \& {Cassinelli}(1987)}]{Wolfire1987ApJ...319..850W}
{Wolfire}, M.~G. \& {Cassinelli}, J.~P. 1987, \apj, 319, 850

\bibitem[{{Wright} {et~al.}(2000){Wright}, {van Dishoeck}, {Black},
  {Feuchtgruber}, {Cernicharo}, {Gonz{\' a}lez-Alfonso}, \& {de
  Graauw}}]{Wright2000A&A...358..689W}
{Wright}, C.~M., {van Dishoeck}, E.~F., {Black}, J.~H., {et~al.} 2000, \aap,
  358, 689

\bibitem[{{Yorke} \& {Sonnhalter}(2002)}]{Yorke2002ApJ...569..846Y}
{Yorke}, H.~W. \& {Sonnhalter}, C. 2002, \apj, 569, 846

\bibitem[{{Zobov} {et~al.}(2000){Zobov}, {Polyansky}, {Tennyson}, {Shirin},
  {Nassar}, {Hirao}, {Imajo}, {Bernath}, \&
  {Wallace}}]{Zobov2000ApJ...530..994Z}
{Zobov}, N.~F., {Polyansky}, O.~L., {Tennyson}, J., {et~al.} 2000, \apj, 530,
  994

\end{thebibliography}

\end{document}